\documentclass[aps,prb,twocolumn,superscriptaddress,amsfonts,amsmath,amssymb]{revtex4}
\usepackage{bm}
\usepackage{graphicx}
\usepackage{amsmath}
\usepackage{amstext}
\usepackage{amssymb}
\usepackage{amsfonts}
\usepackage{amsbsy}
\usepackage{feyn}

\newcommand{\br}{\boldsymbol{r}}
\newcommand{\bq}{\boldsymbol{q}}
\newcommand{\bS}{\boldsymbol{S}}
\newcommand{\bx}{\boldsymbol{x}}
\newcommand{\by}{\boldsymbol{y}}
\newcommand{\bR}{\boldsymbol{R}}
\newcommand{\bQ}{\boldsymbol{Q}}
\newcommand{\bpprop}{\feyn{g}\annotate{-0.5}{-0.2}{\text{\large{O}}}}
\newcommand{\fermiloop}{\feyn{fl}\annotate{-1}{0.56}{\blacktriangleleft}\annotate{-1}{-0.77}{\blacktriangleright}}
\newcommand{\fermiprop}{\feyn{f}\annotate{-0.45}{-0.1}{\blacktriangleright}}
\newcommand{\kslash}{\not \! k}
\newcommand{\qslash}{\not \! q}

\begin{document}

\title{Algebraic spin liquid as the mother of many competing orders}
\author{Michael Hermele}
\affiliation{Department of Physics, University of California, Santa 
Barbara, California 93106}
\author{T. Senthil}
\affiliation{Center for Condensed Matter Theory, Indian Institute of Science, Bangalore, India 560012}
\affiliation{Department of Physics, Massachusetts Institute of 
Technology, Cambridge, Massachusetts 02139}
\author{Matthew P. A. Fisher}
\affiliation{Kavli Institute for Theoretical Physics, University of 
California, Santa Barbara, California 93106}
\date{\today}
\begin{abstract}
We study the properties of a class of two-dimensional interacting critical
states -- dubbed algebraic spin liquids -- that can arise in two-dimensional quantum magnets. 
A particular example that we focus on is 
the staggered flux spin liquid, which plays a key role in some theories of underdoped
cuprate superconductors.  We show that the low-energy theory of such states has much higher symmetry 
than the underlying microscopic spin system.  This symmetry has 
remarkable consequences, leading in particular to the unification of a number 
of seemingly unrelated competing orders.  The correlations of these orders -- including, in the staggered flux state, the N\'{e}el vector and the order parameter for the columnar and box valence-bond solid states -- all exhibit the \emph{same} slow power-law decay.  Implications for experiments in the pseudogap regime of the cuprates and for numerical calculations on model systems are discussed.

\end{abstract}
\maketitle

\section{Introduction}
\label{sec:intro}

In the effort to explain the still-mounting puzzles in many strongly correlated materials, 
one frequently invoked idea is that of competing orders.  
Specifically, it is often appealing to contemplate the presence of rather slowly-varying fluctuations 
in two or more different order parameter degrees of freedom.  
In some cases these orders are not obviously related to one another -- one oft-discussed example is 
antiferromagnetism and superconductivity in the cuprate high-$T_c$ superconductors.\cite{so5}  This kind of situation 
raises an important question:  are the competing orders controlled, all together, by the universal physics 
of a single phase or critical point?  We can also turn this question on its head -- rather than phenomenologically 
introducing some set of slowly-fluctuating orders, we can take a somewhat more microscopic approach and look for possible 
quantum states of a given system.  Then we can ask whether competing orders arise naturally in some such state.

In this paper we shall follow this strategy and show that, somewhat surprisingly, this physics obtains 
within a certain spin liquid state\cite{pwa} of two-dimensional electronic Mott insulators that has been 
suggested to play a key role in the underdoped cuprate 
superconductors.\cite{affleck-marston-rapid,affleck-marston-prb,wen-lee-su2,kim-lee,
rantner-wen-prl,senthil-lee}  
The particular spin liquid  state we consider has been variously described as a ``$d$-wave'' resonating 
valence bond (RVB) state or a staggered flux (sF) state. Here we will use the latter nomenclature and refer to it as the staggered flux state.  It is important to note that the staggered flux spin liquid possesses no broken symmetries and is quite distinct from ordered states with a staggered pattern of orbital currents; instead, it is a specific incarnation of the RVB idea of Anderson.\cite{pwa-triangular}
 
Previous papers have shown that the sF spin liquid is an interacting critical state, and that it may be a stable \emph{critical phase}\cite{rantner-wen-prl,xgw-qorder-ssl,RWspin,stable-u1} --  
the spin correlations decay as a power of the distance 
with a universal exponent, and, while a description in terms of fractional $S = 1/2$ spinons is natural, they do not behave as free quasiparticles even at asymptotically low energy.  Furthermore, the dynamic critical exponent $z = 1$.  Alternatively,  the long-distance, low-energy properties are controlled 
by an interacting, conformally invariant fixed point.
Such states were dubbed algebraic spin liquids (ASL) in Ref.~\onlinecite{rantner-wen-prl}.
Here we show that,   
remarkably, several competing orders are unified within the sF state by an emergent $SU(4)$ symmetry, and all have the 
same slowly-varying long-distance correlations.  

\begin{figure}
\includegraphics[width=6cm]{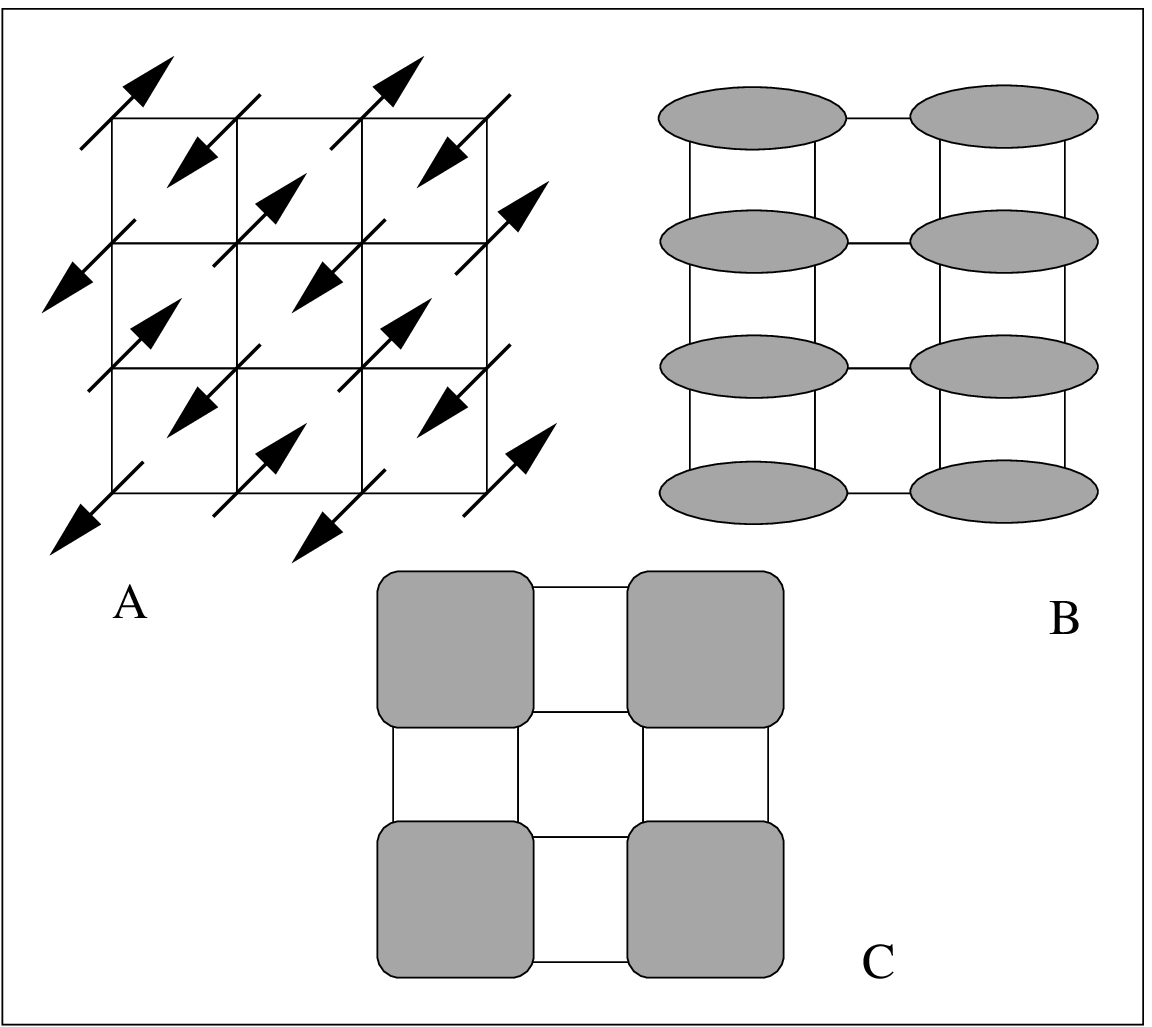}
\caption{Cartoon pictures of some of the slowly-varying competing orders within the staggered flux spin liquid state.  These are the N\'{e}el state (A), and the columnar (B) and box (C) valence bond solids.  The shaded regions denote those groups of spins that are most strongly combined into local singlets.  Note that in the sF state these orders fluctuate in both space and time.  These pictures describe the character of some of the important slowly-varying fluctuations, but should \emph{not} be viewed as snapshots of the physics at the lattice scale, which may be quite complicated.}
\label{fig:orders}
\end{figure}

Two of the competing orders are simply the the N\'{e}el vector, 
and the order parameter for the columnar and box valence-bond solid (VBS) states -- cartoon pictures of these orders are shown in Fig.~\ref{fig:orders}.  To be precise, consider a model Mott insulator on the square lattice with one electron per site at zero temperature, and suppose the system has been tuned into the staggered flux spin liquid by increasing the magnetic frustration.  This state should be present in the parameter space of this system at least as a multicritical point but potentially as a stable phase.  A measurement of the N\'{e}el correlations will find the power-law decay
\begin{equation}
\label{eqn:neel-decay}
(-1)^{(r_x + r_y)}\langle \bS_{\br} \cdot \bS_{0} \rangle \sim 1/|\br|^{(1 + \eta)} \text{.}
\end{equation}
Remarkably the VBS correlations display the same behavior; for example,
\begin{equation}
\label{eqn:vbs-decay}
(-1)^{r_x} \langle (\bS_{\br + \bx} \cdot \bS_{\br}) (\bS_{\bx} \cdot \bS_{0} ) \rangle
	 \sim 1/|\br|^{(1 + \eta)}  \text{.}
\end{equation}
This correlator measures the tendency of the system to order into the columnar dimer pattern shown in Fig.~\ref{fig:orders}B.  Furthermore, several other (more unusual) competing orders with the \emph{same} power-law decay are present.  These are the plaquette-centered spin at $\bq = (\pi, \pi)$, the density of Skyrmions in the N\'{e}el vector at $\bq = (\pi, \pi)$, and a kind of triplet valence bond order that breaks spin rotations but not time reversal.  Also, the uniform spin chirality exhibits slow power-law decay with an exponent that is likely the same as for the other orders.  It is important to note that the decay may be quite slow and thus is potentially observable in experiments and numerical simulations -- variational wavefunction studies\cite{arun-wavefn,ivanov-thesis} of the N\'{e}el correlations provide the rough guess $\eta \approx 0.5$.

This physics may have important consequences for the pseudogap regime of the underdoped cuprates.  
It has been suggested that this part of the phase diagram may be viewed as a doped sF spin liquid.\cite{wen-lee-su2,kim-lee,rantner-wen-prl,senthil-lee}  In particular, in the spin sector, 
the rather high-temperature physics of the pseudogap should be essentially unchanged from the undoped sF state.  
The presence of slowly-fluctuating competing orders related by a large $SU(4)$ symmetry opens up a new possibility for tests of this hypothesis.  
The simplest of these to probe is almost certainly the N\'{e}el vector, the fluctuations of which are directly measured by 
magnetic neutron scattering at $\bq = (\pi,\pi)$.  Furthermore, the sF state is described by a critical theory, so if it is present 
this magnetic scattering should exhibit critical scaling.
If this is found it will be important to think about whether the other competing orders related to the 
N\'{e}el vector by $SU(4)$ symmetry can be directly probed.  
These points are discussed in more detail in Sec.~\ref{sec:cuprates} -- readers not interested in following the more technical aspects of our results may wish to skip to this point.

The sF state is just one member of a class of ASLs that also give rise to a variety of competing 
orders unified by a large emergent symmetry.  
Another state of particular interest is the $\pi$-flux ($\pi$F) spin liquid\cite{affleck-marston-rapid,affleck-marston-prb} 
of an $SU(4)$ Heisenberg-like model on the square lattice.  
Assaad has recently carried out quantum Monte Carlo simulations of this model;\cite{assaad} the results suggest 
that the $\pi$F state may have been observed, but further tests are desirable.  The results of this paper can be 
tested numerically and should significantly aid the resolution of this issue.  We give concrete suggestions 
along these lines in Sec.~\ref{sec:su4-model}.

From a formal point of view, the sF and $\pi$F states can both be described at low energies by a field theory of fermionic spinons with massless 
Dirac dispersion, minimally coupled to a noncompact $U(1)$ gauge field.  This theory is often referred to as 
noncompact ${\rm QED_3}$.  There is good evidence that it can flow to a 
conformally invariant, interacting fixed point, over which one has control in the limit of a large number ($N_f$) of fermion flavors.\cite{appelquist,kogut-numerics}  
This fixed point is the algebraic spin liquid.  Here $N_f$ is the number of two-component Dirac fermion fields, and 
these can be rotated into one another by an $SU(N_f)$ flavor symmetry -- we have $N_f = 4$ for the sF state and $N_f = 8$ for 
the $\pi$F  state of the $SU(4)$ spin model.  It has been shown that, in the large-$N_f$ limit, all relevant  perturbations 
to the $\pi$F state are forbidden by symmetry and it is thus a stable phase.\cite{stable-u1}  The same conclusion is easily seen to hold for the sF state (see Sec.~\ref{sec:stability}).  It is not known whether stability continues to hold for the interesting 
values of $N_f$, although the results of Ref.~\onlinecite{assaad} suggest that the $\pi$F state is stable for $N_f = 8$.  
Even if the sF state is unstable, it should appear as a zero-temperature critical or multicritical point and may still be interesting.

In the field theory, the competing orders discussed above arise as follows:
In the simplest scenario, which is 
suggested by the $1/N_f$ expansion, the dominant correlations are those of an $SU(N_f)$ 
adjoint $N^a$ and a scalar $M$ -- these are bilinears of the fermions.  It is a simple matter to work out how 
these operators transform under the symmetries of the spin model, and to find \emph{symmetry-equivalent} physical 
observables with the same transformation properties; these quantities will all exhibit power-law correlations 
decaying as $1/|\br|^{2 \Delta_N}$ or $1/|\br|^{2 \Delta_M}$, where $\Delta_N, \Delta_M < 2$ are the scaling 
dimensions of $N^a$ and $M$, respectively.  (In fact $\Delta_N = \Delta_M$ to all orders in $1/N_f$, although 
it is not clear whether this holds at finite $N_f$.)  Note that $2 \Delta_N = 1 + \eta$.
Both the N\'{e}el vector and the order parameter for the columnar and box VBS states are symmetry-equivalent to particular components of $N^a$; this gives rise to the power-law decay of Eqs.~(\ref{eqn:neel-decay}) and (\ref{eqn:vbs-decay}).

We note that the structure of competing orders arising from ${\rm QED}_3$ has been discussed previously, from a rather different point of view, in a different physical context.\cite{herbut02, tesanovic02, franz03}  Also, it was recently observed that N\'{e}el and VBS orders can be unified (at the mean-field level) by a chiral rotation at the $\pi$F saddle point.\cite{tanaka05}

We now outline the rest of the paper.
We review the description of the sF spin liquid in Sec.~\ref{sec:description}.  Section~\ref{sec:lattice-continuum} discusses the route from the slave-fermion description of the Heisenberg model to the field theory, and Sec.~\ref{sec:largen} reviews the use of the large-$N_f$ expansion to control the sF fixed point.  In Sec.~\ref{sec:symmetries} we discuss in detail the symmetries of the sF state and their associated conserved currents.  Section~\ref{sec:stability} extends the argument of Ref.~\onlinecite{stable-u1} for the stability of the $\pi$F state at large $N_f$ to the sF state -- the only significant difference is the presence of velocity anisotropy, which is dealt with in Refs.~\onlinecite{vafek-anisotropy} and~\onlinecite{ftv-longpaper}, and Appendix~\ref{app:anisotropy}.  Our main result for the sF state is the identification of the slowly-varying competing orders -- this is discussed in Sec.~\ref{sec:enhanced-bilinears}.  The same is done for some components of the conserved currents in Sec.~\ref{sec:current-observables}.  In Sec.~\ref{sec:su4-model} we shift gears to discuss an analogous identification of competing orders for the $\pi$F state of an $SU(4)$ Heisenberg model.  Finally, in Sec.~\ref{sec:cuprates} we discuss the prospects for observation of this physics in the cuprates, and we conclude in Sec.~\ref{sec:discussion} with a discussion of some of the issues raised by our results.

\section{Describing the Algebraic Spin Liquid}
\label{sec:description}

\subsection{From the lattice to the continuum}
\label{sec:lattice-continuum}

We begin by reviewing the description of the algebraic spin liquid fixed point, and the staggered flux state in particular.\cite{wen-lee-su2, kim-lee, rantner-wen-prl, xgw-qorder-ssl, RWspin, stable-u1,
ftv-longpaper,ft-qed3-prl}  The starting point is the slave-fermion mean-field theory of the $S=1/2$ Heisenberg model on the square lattice:
\begin{equation}
{\cal H} = J \sum_{\langle \br \br' \rangle} \boldsymbol{S}_{\br} \cdot \boldsymbol{S}_{\br'} + \cdots \text{.}
\label{eqn:spin-model}
\end{equation}
Here $J > 0$ (antiferromagnetic exchange), and the ellipsis represents perturbations consistent with the symmetries, such as further neighbor frustrating exchanges, ring exchange terms, and so on.  We require that the Hamiltonian be invariant under $SU(2)$ spin rotations, time reversal, and the full space group of the square lattice.  Formally we may rewrite the spin as a bilinear of fermionic ``spinon'' operators
\begin{equation}
\boldsymbol{S}_{\br} = \frac{1}{2} f^\dagger_{\br \alpha} \boldsymbol{\sigma}_{\alpha \beta}
f^{\vphantom\dagger}_{\br \beta} \text{.}
\end{equation}

Here $\alpha = 1,2$, and $f^\dagger_{\br 1}$ ($f^\dagger_{\br 2}$) creates a spin-up (spin-down) fermion.
This is an exact rewriting when combined with the local constraint $f^\dagger_{\br \alpha} f^{\vphantom\dagger}_{\br \alpha} = 1$.  
Exploiting the well-known $SU(2)$ gauge redundancy in spinon variables,\cite{su2g,su2g-2} the spin-spin interaction is decoupled with an $SU(2)$ gauge field residing on the links of the lattice.  The mean-field saddle points are then described by quadratic spinon Hamiltonians; depending on the structure of the saddle point, the important low-energy fluctuations enter via an $SU(2)$, $U(1)$ or $Z_2$ gauge field minimally coupled to the spinons.\cite{wen-old-mft}

At the mean-field level, the sF state is described by the Hamiltonian
\begin{equation}
{\cal H}^0_{\rm{sF}} = - \sum_{\br \in A} \, \sum_{\br' {\rm n. n.} \br}
\big[ (i t + (-1)^{(r_y - r'_y)}\Delta ) f^\dagger_{\br \alpha} 
  f^{\vphantom\dagger}_{\br' \alpha}
+ \text{H. c.} \big] \text{,}
\label{eqn:mf-hamiltonian}
\end{equation}
where the first sum is over sites in the $A$ sublattice, and the second is over the nearest neighbors of $\br$.  This describes spinons hopping in a background staggered flux of $\Phi = \pm 4\arctan(t/\Delta)$, where the sign alternates from one sublattice of square plaquettes to the other.  The apparent breaking of translation symmetry is a gauge artifact; the spinons transform under certain lattice symmetries with an additional $SU(2)$ gauge transformation.  Physical operators are gauge invariant, so their transformation properties are unaffected and the saddle point possesses the full symmetry of the microscopic model.  This situation is summarized by saying that the spinons obey a \emph{projective symmetry group}\cite{xgw-qorder-ssl} (PSG).  The action of the PSG on the spinons is specified in detail in Appendix~\ref{app:cont-limit}

The low-energy fluctuations about Eq.~(\ref{eqn:mf-hamiltonian}) are encapsulated by a compact $U(1)$ gauge field minimally coupled to the spinons.  
The full lattice Hamiltonian takes the form
\begin{eqnarray}
\label{eqn:full-latt-hamiltonian}
&& {\cal H}_{{\rm sF}} = h \sum_{\langle \br \br' \rangle} e^2_{\br \br'}
- K \sum_{\square} \cos (\operatorname{curl} a) \\
&& - \sum_{\br \in A} \, \sum_{\br' {\rm n. n.} \br}
\big[ (i t + (-1)^{(r_y - r'_y)}\Delta ) f^\dagger_{\br \alpha} 
  e^{-i a_{\br \br'}} f^{\vphantom\dagger}_{\br' \alpha} + \text{H. c.} \big] \text{.} \nonumber
\end{eqnarray}
Here $e_{\br \br'}$ and $a_{\br \br'}$ are lattice vector fields: $e$ is the electric field and takes integer eigenvalues, while $a$, the vector potential, is a $2\pi$-periodic phase.  On the same link of the lattice, $e$ and $a$ satisfy the canonical commutation relation $[a, e] = i$.  The second term of Eq.~(\ref{eqn:full-latt-hamiltonian}) is a sum over square lattice plaquettes, and $(\operatorname{curl} a)$ is the discrete line integral of the vector potential taken counterclockwise around the given plaquette.  The Hamiltonian must be supplemented by the gauge constraint
\begin{equation}
(\operatorname{div} e)_{\br} + f^\dagger_{\br \alpha} f^{\vphantom\dagger}_{\br \alpha} = 1 \text{,}
\end{equation}
where $(\operatorname{div} e)_{\br}$ is the lattice divergence of the electric field.
This gauge theory reduces exactly to the nearest-neighbor Heisenberg model in the limit $K = 0$ and 
$h/t \to \infty$; in this limit $e \equiv 0$ and the gauge constraint becomes $f^\dagger_{r \alpha} f^{\vphantom\dagger}_{r \alpha} = 1$.  
We will be interested instead in a limit where the mean-field theory is manifestly a good starting point and 
is valid up to intermediate length scales, so we consider $K \gg t \gg h$. (In both cases $\Delta \approx t$.)  
The resulting ASL fixed point will control the low-energy physics for some spin Hamiltonians of the 
general form of Eq.~(\ref{eqn:spin-model}).  If the fixed point is stable, no fine-tuning should be necessary 
to access this part of parameter space, but the precise microscopic requirements are unknown.

In the limit of interest, the gauge fluctuations are strongly suppressed by the large Maxwell term; that is, fluctuations in $(\operatorname{curl} a)$ at the scale of the lattice are very small.  We can therefore first write a continuum theory of the long-wavelength, low-energy free-fermion excitations of Eq.~(\ref{eqn:mf-hamiltonian}), and then include the gauge fluctuations.  The technical details are outlined in Appendix~\ref{app:cont-limit}; the resulting low-energy theory consists of four massless two-component Dirac fermions minimally coupled to a \emph{noncompact} $U(1)$ gauge field.  Microscopically, the gauge field is compact, which means physically that instanton configurations (magnetic monopoles) are allowed in the action.  Therefore we should view the noncompact theory as a point in the parameter space of the compact theory where all monopole fugacities have been tuned to zero -- we shall be interested in expanding about this point.  As discussed in Ref.~\onlinecite{stable-u1}, monopoles can (and must) be incorporated as perturbations.  This is greatly aided by the observation that the absence of monopoles is precisely equivalent to the presence of an emergent global $U(1)_{{\rm flux}}$ symmetry corresponding to the conservation of gauge flux, which is only conserved modulo $2\pi$ in the compact theory.\cite{dqcp-science,dqcp-longpaper}

For simplicity of notation, it is convenient to suppress all fermion indices and work with the eight-component object $\Psi$.  We express matrices acting on $\Psi$ as tensor products of the Pauli matrices $\tau^i$, $\mu^i$ and $\sigma^i$.  The $\tau^i$ act within the Dirac space of each two-component fermion, the $\sigma^i$ act on $SU(2)$ spin indices, and the $\mu^i$ connect the two different nodes.
The imaginary-time action can be written $S = \int d^3 x {\cal L}_E$, with
\begin{equation}
\label{eqn:euclidean-lagrangian}
{\cal L}_E =  \bar{\Psi} \Big[ -i \gamma^\mu (\partial_\mu + i a_\mu ) \Big] \Psi  +
 \frac{1}{2 e^2} \sum_{\mu} (\epsilon_{\mu\nu\lambda} \partial_\nu a_\lambda )^2 +
 \cdots \text{,}
\end{equation} 
where $\gamma^\mu = (\tau^3, \tau^2, -\tau^1)$ for $\mu = 0,1,2$, respectively, 
and $\bar{\Psi} \equiv i \Psi^\dagger \tau^3$. The observant reader will notice that we have dropped any explicit velocity anisotropy for the fermions; it is instead grouped with the other perturbations consistent with the microscopic symmetries represented by the ellipsis.  

We defer consideration of the perturbations to the first two terms of Eq.~(\ref{eqn:euclidean-lagrangian}) until Sec.~\ref{sec:stability}.  For now we simply drop them.  It is immediately clear that the resulting theory has a much higher symmetry than that of the spin model.  In addition to the $U(1)_{{\rm flux}}$ symmetry discussed above, there is an $SU(4)$ flavor symmetry acting on the fermions.  This symmetry is generated by the $4 \times 4$ traceless, Hermitian matrices $T^a$, where $a = 1,\dots,15$.  The $T^a$ can be expressed in terms of tensor products of the $\sigma^i$ and $\mu^i$ Pauli matrices; that is, they are linear combinations of the basis $\{ \sigma^i, \mu^i, \sigma^i \mu^j \}$.
The action on the fermion fields is given by:
\begin{eqnarray}
\Psi &\to& \exp(i \lambda^a T^a) \Psi \\
\bar{\Psi} &\to& \bar{\Psi} \exp(-i \lambda^a T^a) \nonumber \text{.}
\end{eqnarray}
Note that this is a \emph{flavor} symmetry; that is, it rotates the 4 two-component fermions into one another, 
but does not affect the Dirac structure.  More precisely, $[T^a, \gamma_\mu] = 0$.  The remarkable 
consequences of this $SU(4)$ symmetry are the main focus of this paper.

\subsection{The large-$N_f$ limit}
\label{sec:largen}

The field theory of Eq.~(\ref{eqn:euclidean-lagrangian}) has a non-trivial conformally invariant fixed point that is not amenable to a direct analytical treatment.  As with other 
such critical theories in $2+1$ dimensions, the best that can be done is to 
deform the model to a limit where we do have control and use this to understand as much as possible about 
the case of physical interest. A useful and familiar analogy is with the critical fixed point of the classical 
$O(n)$ model in three dimensions. To access this fixed point analytically it is necessary to study it in an 
$\epsilon$-expansion near four dimensions or in an expansion in $1/n$ directly in three dimensions. 
Both of these expansions similarly provide useful analytic access in the present problem as well. However, the 
fixed point that describes the theory in Eq.~(\ref{eqn:euclidean-lagrangian}) has no relevant perturbations [in contrast to the $O(n)$ critical fixed point]. Here we will 
we follow previous works\cite{appelquist,kim-lee,rantner-wen-prl,RWspin,ft-qed3-prl,stable-u1,kaveh04} and
generalize the theory to a large number of fermion
flavors by adding an extra index to the Dirac field: $\Psi \to \Psi_a$, where $a = 1, \dots, N_f/4$.  
With this convention $N_f$ is the number of two-component Dirac fermions and the flavor symmetry is 
enlarged to $SU(N_f)$.  $N_f=4$ corresponds to the physical case of $SU(2)$ spin.  For $N_f$ sufficiently large, 
it is reasonable to treat $1/N_f$ as a formal expansion parameter.  Provided we take $e^2 \sim 1/N_f$, and 
in the absence of perturbations, the theory can be solved order-by-order in $1/N_f$.  This can be carried 
out simply in terms of diagrams, and is described in Appendix~\ref{app:largen-diagrams}.

It is believed that the large-$N_f$ expansion describes a conformally invariant fixed point to all orders in $1/N_f$.\cite{appelquist}  This fixed point is the algebraic spin liquid.  At $N_f = \infty$ the theory is scale-invariant and the fermions behave for most purposes as if they were free.  (That the fermions are not truly free is apparent from the presence of operators that acquire an anomalous dimension even at $N_f = \infty$; see the discussion of the gauge charge current in Sec.~\ref{sec:symmetries}.  This is analogous to the situation with the quadratic ``mass'' operator in the $O(n)$ model.)  The $1/N_f$ corrections to this extreme limit correspond physically to incorporating gauge fluctuations, and one finds that various correlators acquire anomalous dimensions for $N_f < \infty$.  The usual justification for the presence of a conformally invariant fixed point comes from a consideration of the above perturbation theory for an arbitrary correlation function.  Due to the $1/|q|$ form of the photon propagator, the effective expansion parameter for this perturbation theory is easily seen to be dimensionless.  
Because there is no longer any scale in the problem (aside from a short-distance cutoff) it is natural to expect that the large-$N_f$ expansion describes a conformally invariant fixed point.  Furthermore, a fermion mass cannot be generated perturbatively in $1/N_f$ because all such terms break either the $SU(N_f)$ flavor symmetry, or parity and time reversal.

Further insight is provided by a renormalization group (RG) approach perturbative in $1/N_f$, which is also very useful as a tool for calculation.\cite{RWspin, vafek-anisotropy, ftv-longpaper, stable-u1,kaveh04}  One simply regards $1/N_f$ as an exactly marginal perturbation to the $N_f = \infty$ fixed point, and calculates corrections to the properties of this fixed point as an asymptotic series in $1/N_f$.  For technical purposes it is most convenient to implement a ``field theory'' RG, and first calculate some correlation function to the desired order in $1/N_f$ with a fixed UV cutoff.  Then we demand that this correlator satisfy the appropriate Callan-Symanzik equation, which is simply the mathematical statement that we can equivalently change the overall momentum scale $k \to e^{-\ell} k$, or rescale the fields.  We can also include perturbations to the fixed point; then we must also rescale their coupling constants, and the resulting Callan-Symanzik equations allow us to calculate the flow equations for the couplings.

In this language, the statement that the $1/N_f$ expansion describes a scale-invariant fixed point to all orders can be put as follows:  Set all perturbations to the fixed-point theory to zero.  Then we can write a Callan-Symanzik equation for any correlator involving only the anomalous dimensions of fields.  If these equations can all be satisfied order-by-order in $1/N_f$, then the theory is indeed scale-invariant.  While it has not been proven that this is the case, to our knowledge no inconsistency has been found.

\section{Symmetries and Stability of the Spin Liquid}

\subsection{Symmetries and conserved currents}
\label{sec:symmetries}

The above considerations strongly suggest that for sufficiently large but finite $N_f$, somewhere in its parameter space 
the field theory has a conformally invariant fixed point smoothly connected to the $N_f = \infty$ fixed point.  
It is reasonable, although not certain, that this fixed point continues to exist for $N_f=4$, and that a 
qualitative picture of its properties is provided by low-order calculations in the 
$1/N_f$ expansion.  This should be viewed as providing a definition of the algebraic spin liquid fixed point.  

The ASL then has the symmetry group
\begin{eqnarray}
\label{eqn:asl-sg}
SG_{{\rm ASL}} &=& (\text{Conformal Symmetry}) \\ 
&\times& {\cal C} \times {\cal P} \times {\cal T}
\times SU(4)_{{\rm flavor}} \times U(1)_{{\rm flux}} \nonumber \text{.}
\end{eqnarray}
This is a much larger symmetry than is present in the microscopic model.
In addition to the $SU(4)_{{\rm flavor}}$, $U(1)_{{\rm flux}}$, and conformal symmetries discussed above, 
we also have the discrete symmetries of charge conjugation (${\cal C}$), parity (${\cal P}$), and 
time reversal (${\cal T}$).  It is important to note that here we are referring to the 
symmetries of the continuum theory, and not, for example, to the time-reversal symmetry of the spin model.  
The action of this operation on the field theory degrees of freedom will involve the continuum ${\cal T}$ 
combined with other operations.  

In Sec.~\ref{sec:stability} we show that, in the large-$N_f$ limit, 
all allowed perturbations are irrelevant and the staggered flux ASL is thus a stable phase.  For $N_f=4$ 
it is then likely that the sF state either remains stable or has only a small number of unstable directions.  
Since the physical case of $N_f=4$ is of primary interest, we focus here on this case but use the large-$N_f$ expansion to control our results.

We wish to put special emphasis on the $SU(4)$ flavor symmetry discussed in 
Sec.~\ref{sec:lattice-continuum}, 
which leads to a host of heretofore unnoticed consequences.  In order to understand the relationship to 
the microscopic symmetries, it is useful to think in terms of the subgroup
\begin{equation}
SU(2)_{{\rm spin}} \times SU(2)_{{\rm nodal}} \subset SU(4) \text{,}
\end{equation}
where $SU(2)_{{\rm spin}}$ is the physical spin and is generated by $\sigma^i$. $SU(2)_{{\rm nodal}}$ 
consists of emergent symmetries and is generated by the Pauli matrices $\mu^i$.  We refer to this as 
the ``nodal'' $SU(2)$ because these flavor rotations involve the two distinct nodes arising from 
the staggered flux band structure and commute with $SU(2)$ spin rotations.  Certain discrete $SU(2)_{{\rm nodal}}$ 
rotations are intimately tied to the microscopic lattice symmetries; this is apparent upon inspection of 
the symmetry transformation laws enumerated in Appendix~\ref{app:cont-limit}.

The conserved $SU(4)$ flavor current is
\begin{equation}
J^a_\mu = -i \bar{\Psi} \gamma_\mu T^a \Psi \text{.}
\end{equation}
This multiplet of operators transforms as an $SU(4)$ adjoint and a Lorentz vector.  
It is easy to show by explicit calculation in the $N_f = \infty$ theory that the two-point 
function of $J^a_\mu$ falls off as $1/x^4$, and $J^a_\mu$ thus has dimension 2.  Since 
conserved currents cannot acquire an anomalous dimension, this must hold for all $N_f$.

The emergent $U(1)_{{\rm flux}}$ symmetry also plays an important role.  
This symmetry is associated with the conserved gauge flux current
\begin{equation}
\label{eqn:gauge-flux-current}
j^G_\mu = \epsilon_{\mu \nu \lambda} \partial_\nu a_\lambda \text{.}
\end{equation}
This current also has scaling dimension 2 for all $N_f$.
The conservation law $\partial_\mu j^G_\mu = 0$ is violated precisely by magnetic monopoles, 
so the emergent $U(1)_{{\rm flux}}$ symmetry encapsulates the irrelevance of monopoles at low energies.\cite{dqcp-science,dqcp-longpaper}  
Monopole operators are those carrying a nonzero $U(1)_{{\rm flux}}$ charge.  

There is also a conserved $U(1)$ gauge charge current $G_\mu = -i \bar{\Psi} \gamma_\mu \Psi$ 
associated with the global gauge transformation $\Psi \to e^{i \phi} \Psi$.  Since this gauge ``symmetry'' 
is not a true symmetry but rather the consequence of a redundancy in our choice of variables, it should 
not be surprising that $G_\mu$ is rather special.  Making the infinitesimal change of variables
$a_\mu (x) \to a_\mu (x) + \epsilon_\mu (x)$ in the functional integral leads via standard manipulations to 
the Maxwell equation, which we can regard as an operator identity:
\begin{equation}
\label{eqn:g-mu-op-identity}
G_\mu = \frac{i}{e^2} \epsilon_{\mu \nu \lambda} \partial_\nu j^G_\lambda + (\text{more irrelevant terms}) \text{.}
\end{equation}
As with all operator identities, the meaning of this equation is that in any correlation function 
$G_\mu(x)$ can be replaced by the right-hand side of Eq.~(\ref{eqn:g-mu-op-identity}), as long as 
the other fields involved in the correlator are not close to $x$.  This implies that $G_\mu$ has dimension 3, 
as can be verified by explicit calculation in the $N_f = \infty$ theory.  Furthermore, $G_\mu$ should be 
thought of as a derivative of $j^G_\mu$ and not a new truly independent operator.  We note that this is a manifestation of the fact that the fermions are not quite free, even at $N_f = \infty$.

\subsection{Stability of the spin liquid}  
\label{sec:stability}

In order to assess the stability of the sF state at large $N_f$, we need only follow the 
argument of Ref.~\onlinecite{stable-u1} for the stability of the $\pi$-flux ($\pi$F) state of an $SU(N)$ magnet.  
The field theories in these two cases are identical, but different perturbations are allowed by 
the microscopic symmetries.  As in Ref.~\onlinecite{stable-u1}, we can group all 
operators at the sF algebraic spin liquid fixed point into two classes: those carrying $U(1)_{{\rm flux}}$ charge 
and those that do not.  The first class is comprised of the monopole operators, which are all strongly irrelevant at large $N_f$, with scaling dimensions proportional to $N_f$.\cite{bkw}

The monopole-free sector of the theory must be considered in more detail, as it contains 
operators that would be relevant if allowed by symmetry.  As in Ref.~\onlinecite{stable-u1}, 
the potentially dangerous perturbations are fermion bilinears with zero and one 
derivative (``mass'' and ``kinetic'' terms, respectively). It is a simple exercise to 
show that all mass terms are forbidden by symmetry.  In fact, this is true even if one 
only considers $SU(2)$ spin rotations, time reversal, and $x$- and $y$- translations.

We must also consider the kinetic terms as these have dimension $3 + {\cal O}(1/N_f)$ and are 
exactly marginal at infinite $N_f$.
Two such terms are allowed by symmetry.  The first is simply 
the isotropic kinetic energy
\begin{equation}
K_s = -i \Psi^\dagger \Big[ \tau^1 (\partial_1 + i a_1) + \tau^2 (\partial_2 + i a_2) \Big] \Psi \text{.}
\end{equation}
This term has no effect, as it can be absorbed into the fixed point 
theory Eq.~(\ref{eqn:euclidean-lagrangian}) by rescaling the time coordinate.  

The second term is a velocity anisotropy for the fermions and cannot be 
removed by rescaling space and time.  It is therefore important to know 
the $1/N_f$ correction to the scaling dimension of this operator, which takes 
the form
\begin{equation}
K_a = -i \delta \Psi^\dagger \mu^3 \Big[ \tau^1 (\partial_1 + i a_1) - \tau^2 ( \partial_2 + i a_2) \Big] 
\Psi^{\vphantom\dagger} \text{,}
\end{equation}
where it should be noted that the coefficient is $\delta$.
The RG flow of $\delta$ can be calculated as a function of $1/N_f$ and $\delta$.  
This was done by Vafek, Tesanovic, and Franz\cite{vafek-anisotropy,ftv-longpaper} to 
leading order in $1/N_f$ and to all orders in $\delta$; a more straightforward but 
equivalent calculation is discussed in Appendix~\ref{app:anisotropy} and reproduces 
the leading term of their result.  We find
\begin{equation}
\frac{d \delta}{d \ell} = - \frac{64}{5 \pi^2 N_f} \delta + {\cal O}(\delta^2/N_f , \delta/N_f^2) \text{,}
\end{equation}
and the velocity anisotropy is therefore irrelevant for sufficiently large $N_f$.

Other gauge-invariant operators in the monopole-free sector of the theory can be 
constructed by forming polynomials of the fermion fields and the gauge flux, and 
inserting covariant derivatives.  It is easy to see that all such operators are trivially 
irrelevant in the $N_f \to \infty$ limit, and thus do not destabilize the algebraic spin 
liquid at sufficiently large $N_f$.

\section{Physical Observables}

In this section we examine two classes of field theory operators, focusing on their connection to 
observables in the spin model.  The first class has slowly-varying correlations and gives rise to 
the competing orders within the algebraic spin liquid.  The second class is comprised of 
the conserved currents $J^a_\mu$ and $j^G_\mu$.

Quite generally, any operator in the spin model is connected to the field theory by the relation
\begin{equation}
\label{eqn:eft-reln}
{\cal O}_{{\rm Spin\, Model}} \sim \sum_i c_i {\cal O}^i_{{\rm Field\, Theory}} \text{.}
\end{equation}
The meaning of this expression is that the long-distance correlations of the spin model operator are identical to those of the sum of field theory operators on the right-hand side.
The $c_i$ are nonuniversal coefficients, and generically $c_i \neq 0$ if and only if
${\cal O}^i_{{\rm Field\, Theory}}$ transforms identically to the spin model operator 
under the \emph{microscopic} symmetries -- when this is the case we say the two operators 
are \emph{symmetry-equivalent}.  More precisely, all terms on both sides of Eq.~(\ref{eqn:eft-reln}) 
should transform in the same irreducible representation of the microscopic symmetry group, and all 
field theory operators transforming in this representation will contribute.

\subsection{Fermion bilinears with enhanced correlations}
\label{sec:enhanced-bilinears}

Here we shall be interested in the $SU(4)$ adjoint
\begin{equation}
N^a = -i \bar{\Psi} T^a \Psi \text{,}
\end{equation}
as well as the scalar
\begin{equation}
M = -i \bar{\Psi} \Psi \text{.}
\end{equation}
Rantner and Wen calculated the two-point function, and hence the scaling dimension, of 
one member of the $N^a$ multiplet to leading order in the $1/N_f$ expansion\cite{RWspin} -- 
the operator they considered is symmetry-equivalent to the N\'{e}el vector.  By $SU(4)$ symmetry their result 
applies to the entire multiplet and its scaling dimension is
\begin{equation}
\Delta_N  = 2 - \frac{64}{3 \pi^2 N_f} + {\cal O}(1/N_f^2) \text{.}
\end{equation} 
The correlations of these operators are therefore \emph{enhanced} by gauge fluctuations -- 
this is physically very reasonable, since the gauge force tends to bind the oppositely 
charged $\Psi$ and $\bar{\Psi}$ particles.

Although they are not related by any obvious symmetry, $M$ and $N^a$ have the same scaling 
dimension to all orders in $1/N_f$; this is shown in Appendix~\ref{app:allorders}.  At present it is 
not clear if there is a deeper reason (\emph{e.g.} some hidden symmetry) for this equality.  Therefore, 
while this statement may hold at finite $N_f$, it may also be merely an accident of the large-$N_f$ 
expansion that is destroyed by effects nonperturbative in $1/N_f$. Whether or not $M$ and $N^a$ 
have the same scaling dimension, the large-$N_f$ expansion indicates at least that the 
correlations of $M$ are also enhanced by gauge fluctuations.

At large $N_f$, $M$ and $N^a$ are the most relevant operators and give the dominant long-distance correlations.  
They are also in some sense the most natural instabilities of the algebraic spin liquid, although a proper 
treatment of this issue requires consideration of the parameter space around the fixed point and 
not only the ASL itself.  

It is useful to find physical observables symmetry-equivalent to $N^a$ and $M$ as their 
correlations may decay slowly enough to be readily observable.  This exercise is easily 
carried out making use of the transformation laws of Appendix~\ref{app:cont-limit}; here we summarize 
and discuss the results.
It is convenient to group the $N^a$ into three classes depending on their matrix structure:
\begin{eqnarray}
\boldsymbol{N}^i_A &=& -i \bar{\Psi} \mu^i \boldsymbol{\sigma} \Psi \text{,} \\
\boldsymbol{N}_B &=& -i \bar{\Psi} \boldsymbol{\sigma} \Psi \text{,} \\
N^i_C &=& -i \bar{\Psi} \mu^i \Psi \text{.}
\end{eqnarray} 
Symmetry-equivalent spin operators are listed in Table~\ref{tab:enhanced-ops}.  

\begin{table}
\begin{ruledtabular}
\begin{tabular}{c | c }
	$\text{  Field Theory  }$ & $\text{  Spin Model  }$ \\
	\hline
	$\boldsymbol{N}^1_A \text{ , } \boldsymbol{N}^2_A$ &
	$(-1)^{r_x + 1} \boldsymbol{S}_{\br} \times \boldsymbol{S}_{\br + \boldsymbol{y}} \text{ , }
	(-1)^{r_y} \boldsymbol{S}_{\br} \times \boldsymbol{S}_{\br + \boldsymbol{x}}$ \\
	\hline
	$\boldsymbol{N}^3_A$ & $(-1)^{r_x + r_y} \boldsymbol{S}_{\br}$ \\	
	\hline
	$\boldsymbol{N}_B$ & $(-1)^{r_x + r_y} \big[ (\bS_1 + \bS_3)(\bS_2 \cdot \bS_4)$ \\
		& $\qquad\qquad + (\bS_2 + \bS_4) (\bS_1 \cdot \bS_3) \big]$ \\
	\hline
	$N^1_C \text{ , } N^2_C$ & $(-1)^{r_y} \bS_{\br} \cdot \bS_{\br + \by} \text{ , }
						 (-1)^{r_x} \bS_{\br} \cdot \bS_{\br + \bx}$ \\
	\hline
	$N^3_C$ & $\big[ \bS_1 \cdot (\bS_2 \times \bS_4) - \bS_2 \cdot (\bS_3 \times \bS_1)$ \\
	& $ \,\,\,\,\,\,\,\,\,\,\, + \, \bS_3 \cdot (\bS_4 \times \bS_2) - \bS_4 \cdot (\bS_1 \times \bS_3) \big]$ \\
	\hline
	$M$ & $\big[ \bS_1 \cdot (\bS_2 \times \bS_4) + \bS_2 \cdot (\bS_3 \times \bS_1)$ \\
	& $ \,\,\,\,\,\,\,\,\,\,\, + \, \bS_3 \cdot (\bS_4 \times \bS_2) + \bS_4 \cdot (\bS_1 \times \bS_3) \big]$
\end{tabular}
\end{ruledtabular}
\caption{\label{tab:enhanced-ops} List of observables in the spin model that are symmetry-equivalent to the $N^a$ and $M$ fermion bilinears.
For some of these we label the sites around the plaquette with lower-left corner at $\br$ by the numbers $1,\dots,4$.  Precisely, $\bS_1 = \bS_{\br}$, $\bS_2 = \bS_{\br + \bx}$, $\bS_3 = \bS_{\br + \bx + \by}$ and $\bS_4 = \bS_{\br + \by}$.
}
\end{table}

These operators are order parameters for several slowly fluctuating competing orders.  
It is remarkable that these apparently unrelated fluctuations are perfectly balanced within the algebraic spin liquid.  
Two of these observables are quite familiar: $\boldsymbol{N}^3_A$ is the N\'{e}el vector, and 
$\vec{\Phi}_{{\rm VBS}} = (N^2_c, N^1_c)$ is the valence-bond solid order parameter.  
When $\vec{\Phi}_{{\rm VBS}}$ has an expectation value along  the $x$ or $y$ axis the 
columnar VBS state (Fig.~\ref{fig:orders}B) results, while if it points at $45^\circ$ from the axes the resulting state 
is the box VBS (Fig.~\ref{fig:orders}C).  We note that this unification of N\'{e}el and VBS order is quite different from the 
situation at the recently elucidated quantum critical point between N\'{e}el and VBS states,\cite{dqcp-science,dqcp-longpaper} 
where the order parameters have different scaling dimensions [in the case with full $SU(2)$ spin symmetry]
and are not related by symmetry.

The other order parameters are rather unusual and correspond to more exotic ordered states.   
$\boldsymbol{N}_B$ transforms like a plaquette-centered spin at $\boldsymbol{q} = (\pi,\pi)$.  
The operators $\boldsymbol{N}^{1,2}_A$ form the order parameter for a kind of \emph{triplet} 
valence bond solid that breaks spin rotations but not time reversal.  

$N^3_C$ is somewhat more familiar.  If we denote the slowly-varying N\'{e}el field by the 
unit vector field $\boldsymbol{n}$, we can define the density of the familiar topological 
Skyrmion configurations by writing
\begin{equation}
\rho_S = \frac{1}{4\pi} \boldsymbol{n} \cdot ( \frac{\partial \boldsymbol{n}}{\partial x} \times
\frac{\partial \boldsymbol{n}}{\partial y} ) \text{.}
\end{equation}
As usual, if we take periodic boundary conditions in space, $\int d^2 \br \rho_S$ is an 
integer that changes from one time slice to the next upon encountering an instanton defect 
in the N\'{e}el field (a hedgehog).  Now, $N^3_C$ and the corresponding symmetry-equivalent spin operator 
transform like the $\bq = (\pi, \pi)$ component of $\rho_S$.

Finally we turn to $M$, which is odd under time reversal and reflections, and does not 
transform under any of the other microscopic symmetries; it is thus symmetry-equivalent to the uniform component of the scalar spin chirality.  Furthermore, under the symmetries 
of the ASL fixed point, $M$ transforms exactly like a Chern-Simons term for the gauge 
field $a_\mu$, and if $M$ is added to the Lagrangian (via spontaneous symmetry breaking), a 
Chern-Simons term will also be generated.  The resulting state is a chiral spin liquid supporting gapped spinons with fractional statistics.\cite{kalmeyer-laughlin-prl,kalmeyer-laughlin-prb,wwz}

\subsection{Observables for the conserved currents}
\label{sec:current-observables}

Now we shall discuss symmetry-equivalent spin operators for the two conserved currents $J^a_\mu$ and $j^G_\mu$.  
This is most interesting for the information obtained about the general structure of the sF state.  
As discussed in Sec.~\ref{sec:symmetries}, the corresponding correlations decay as $1/r^4$ and may 
be rather difficult to observe.  Furthermore, we will see below that a given microscopic operator may 
be symmetry-equivalent both to a member of one of the current multiplets \emph{and} another, more relevant 
operator.  In order to predict that a particular correlation function of the microscopic model decays 
as $1/r^4$ we need to know that this does not happen.  For numerical simulations and experiments, 
it is undoubtedly better to begin by looking for the stronger correlations discussed above, and to 
consider the ``fine structure'' of the conserved currents only as a second step.

Rather than systematically considering every component of the conserved currents, 
we only highlight some of the most interesting cases.  We begin with the gauge flux 
current $j^G_\mu$.  In the (continuum) Hamiltonian its components correspond to the 
magnetic flux $\Phi_{{\rm B}} = j^G_0$ and the electric field $E_i = i \epsilon_{i j} j^G_j$.  
We find that the electric field is symmetry-equivalent to a staggered scalar spin chirality, 
taken along \emph{lines} of three adjacent lattice sites:
\begin{eqnarray}
(-1)^{(r_x+r_y)} \bS_{\br - \bx} \cdot (\bS_{\br} \times \bS_{\br + \bx}) &\sim& E_x \text{,}\\
-(-1)^{(r_x + r_y)} \bS_{\br - \by} \cdot (\bS_{\br} \times \bS_{\br + \by}) &\sim& E_y \text{.}
\end{eqnarray}
Labeling as above the four sites of the plaquette with lower-left corner $\br$ by the 
numbers $1,\dots,4$, two operators symmetry-equivalent to the magnetic flux are
\begin{eqnarray}
(-1)^{(r_x + r_y)} &\Big(& (\bS_1 \cdot \bS_2) (\bS_3 \cdot \bS_4) \nonumber \\  
&+& (\bS_2 \cdot \bS_3) (\bS_1 \cdot \bS_4) \Big) \sim \Phi_{{\rm B}}
\label{eqn:mag-flux-1}
\end{eqnarray}
and
\begin{equation}
\label{eqn:mag-flux-2}
(-1)^{(r_x + r_y)} (\bS_1 \cdot \bS_3) (\bS_2 \cdot \bS_4) \sim \Phi_{{\rm B}} \text{.}
\end{equation}
Subtracting Eqs.~(\ref{eqn:mag-flux-1}) and (\ref{eqn:mag-flux-2}) we also see that 
the $\bq = (\pi,\pi)$ component of the usual ring exchange operator (\emph{i.e.} that 
obtained from the Hubbard model at order $t^4/U^3$) is symmetry-equivalent to $\Phi_{{\rm B}}$.

Now we move on to the $SU(4)$ flavor current $J^a_\mu$.
The spin at $\bq = (\pi,0)$ and $\bq = (0,\pi)$ is symmetry-equivalent to two spatial components of the current:
\begin{eqnarray}
(-1)^{r_x} \bS_{\br} &\sim& 
	\Psi^\dagger (\tau^1 + \tau^2) \mu^2 \boldsymbol{\sigma} \Psi^{\vphantom\dagger} \text{,} \\
(-1)^{r_y} \bS_{\br} &\sim&
	\Psi^\dagger (-\tau^1 +\tau^2) \mu^1 \boldsymbol{\sigma} \Psi^{\vphantom\dagger} \text{.}
\end{eqnarray}
Rantner and Wen calculated the leading $1/N_f$ corrections to the correlations of these quantities and 
found no anomalous dimension.\cite{RWspin}  This result is explained by $SU(4)$ symmetry 
(and conformal invariance), which implies these operators have dimension 2 to all orders in 
$1/N_f$ (see Sec.~\ref{sec:symmetries}).

The two components of the VBS order, which already made an appearance above in the $N^a$ multiplet, 
are also symmetry-equivalent to two of the $SU(4)$ conserved densities:
\begin{eqnarray}
(-1)^{r_x} \bS_{\br} \cdot \bS_{\br + \bx} &\sim& \Psi^\dagger \mu^1 \Psi^{\vphantom\dagger} \text{,} \\
(-1)^{r_y}\bS_{\br} \cdot \bS_{\br + \by} &\sim& \Psi^\dagger \mu^2 \Psi^{\vphantom\dagger} \text{.}
\end{eqnarray}
Note that there is no inconsistency in the fact that the VBS order appears in two distinct 
multiplets of the field theory.  This simply means that both field theory operators contribute 
to its long-distance correlations; that is,
\begin{equation}
(-1)^{r_x} \bS_{\br} \cdot \bS_{\br + \bx} \sim
	c_1 \Psi^\dagger \tau^3 \mu^2 \Psi^{\vphantom\dagger} + 
c_2 \Psi^\dagger \mu^1 \Psi^{\vphantom\dagger} + \cdots \text{,}
\end{equation}
where $c_{1,2}$ are nonuniversal constants as in Eq.~(\ref{eqn:eft-reln}).  
This is an example where it is clearly \emph{not} true that VBS correlations fall 
off as $1/r^4$ simply because the VBS order parameter appears in $J^a_\mu$.  This possibility 
must be contemplated for other operators symmetry-equivalent to conserved currents, and in general 
it is necessary to consider symmetry-equivalent field theory operators beyond the fermion bilinears.  
In particular, monopole operators carry nontrivial quantum numbers and may play into these 
considerations.\cite{bkw,monopole-qn}

The Skyrmion density $\rho_S$ is symmetry-equivalent to one of the conserved densities:
\begin{equation}
	\rho_S \sim \Psi^\dagger \mu^3 \Psi^{\vphantom\dagger} \text{.}
\end{equation}
It is very interesting that $\rho_S$ is also conserved at the deconfined critical point between 
the N\'{e}el and VBS states, where it corresponds to the magnetic flux of an emergent gauge field.  
We let $S_\mu$ be the gauge flux current as defined in Ref~\onlinecite{dqcp-longpaper}.  There it is 
denoted as $j^G_\mu$; here we call it $S_\mu$ to emphasize that it is quite distinct 
from the $j^G_\mu$ defined in Eq.~(\ref{eqn:gauge-flux-current}) -- in particular, the 
two currents are not symmetry-equivalent.  We do find, however, that $S_\mu$ \emph{is} 
symmetry-equivalent to the following components of the $SU(4)$ flavor current:
\begin{eqnarray}
S_0 &\sim& \Psi^\dagger \mu^3 \Psi^{\vphantom\dagger} \text{,} \\
S_1 &\sim& \frac{i}{\sqrt{2}} \Psi^\dagger (\tau^1 - \tau^2) \mu^3 \Psi^{\vphantom\dagger} \text{,} \\
S_2 &\sim& \frac{i}{\sqrt{2}} \Psi^\dagger (\tau^1 + \tau^2) \mu^3 \Psi^{\vphantom\dagger} \text{.}
\end{eqnarray}
The presence of a $45^\circ$ rotation in the $\tau$-matrix structure is to be expected, since the 
continuum coordinates of the sF state are rotated from the lattice axes by $45^\circ$ 
(see Appendix~\ref{app:cont-limit}).  This is not the case for the continuum theory of 
Ref.~\onlinecite{dqcp-longpaper}.

Remarkably, then, the conserved Skyrmion current that plays such a key role at 
the N\'{e}el-VBS critical point is also a conserved current of the sF algebraic spin liquid.  
Furthermore, it is contained within the larger structure of the $SU(4)$ flavor symmetry.  
It would be interesting to see if there is a natural route between these two fixed points -- 
the $U(1)_{{\rm flux}}$ symmetry would have to be broken by monopole proliferation, and 
the $SU(4)_{{\rm flavor}}$ would need to be broken down to $SU(2)_{{\rm spin}} \times U(1)$, 
where the $U(1)$ corresponds to Skyrmion number conservation at low energies.  On an even 
more speculative note, perhaps other interesting fixed points, so far undiscovered, also have 
the seeds of their structure hidden within the algebraic spin liquid.

\section{Consequences for the $\pi$-flux State of an $SU(4)$ Heisenberg Model}
\label{sec:su4-model}

Recently, Assaad has carried out a quantum  Monte Carlo study of an $SU(4)$ Heisenberg 
antiferromagnet on the square lattice.\cite{assaad} The results available to date are consistent 
with observation of the $\pi$-flux algebraic spin liquid, first studied in the large-$N_f$ limit by 
Affleck and Marston.\cite{affleck-marston-rapid,affleck-marston-prb} 
This state has a very similar structure to the sF ASL considered up to now in this paper -- 
the primary difference is that now we have a microscopic $SU(4)$ spin symmetry, and 
the microscopic lattice symmetries act differently on the continuum Dirac fields.  
Here there is an emergent $SU(8)$ flavor symmetry, and there are 64 fermion bilinears 
which have correlations enhanced by gauge fluctuations.  One of these is the $SU(4)$ analog 
of the N\'{e}el vector -- Ref.~\onlinecite{assaad} found that its correlations fall off very slowly, as 
$1/r^\alpha$, where $\alpha \approx 1.1 - 1.2$.  

The $SU(8)$ symmetry allows us to make the highly 
nontrivial prediction that certain other observables should have the same long-distance correlations.  This will hold in the $\pi$F state provided that the most relevant operators for $N_f = 8$ are indeed the fermion bilinears discussed below, as is suggested by the large-$N_f$ expansion.  If some other multiplet of operators is more relevant, a similar set of predictions will hold, but for different observables.  It is particularly important to consider monopole operators in this context, since they carry nontrivial quantum numbers\cite{bkw} and may have relatively low scaling dimension; this issue will be considered in more detail in a forthcoming paper.\cite{monopole-qn}
With this one caveat in mind, the results of this paper can be tested numerically, and it should thus be possible to 
determine rather conclusively whether or not the $\pi$F state has indeed been observed in the model of Ref.~\onlinecite{assaad}.

The model is defined in terms of the slave fermions $f_{\br \alpha}$, where 
$\alpha = 1,\dots,4$, and we choose the local constraint $f^\dagger_{\br \alpha} f^{\vphantom\dagger}_{\br \alpha} = 2$.  
Here the $SU(4)$ spin rotations are a microscopic symmetry, again generated by the $4 \times 4$ 
matrices $T^a$, with $a = 1,\dots,15$.  The action of an $SU(4)$ spin rotation on the fermions is
$f_{\br \alpha} \to [ \exp (i \lambda^a T^a)]_{\alpha \beta} f_{\br \beta}$.  
We define the Hermitian $SU(4)$ spin operator
\begin{equation}
S^a_{\br} = f^{\dagger}_{\br \alpha} T^a_{\alpha \beta} f^{\vphantom\dagger}_{\br \beta} \text{,}
\end{equation}
and the Hamiltonian takes the form
\begin{equation}
{\cal H}_{SU(4)} = J \sum_{\langle \br \br' \rangle} S^a_{\br} S^a_{\br'} \text{,}
\label{eqn:su4-hamiltonian}
\end{equation}
with $J > 0$.  

The quartic spin-spin interaction can be decoupled with a compact $U(1)$ gauge field, and 
one can consider a mean-field theory (exact at infinite $N_f$) where this gauge field becomes a 
non-fluctuating classical background.  In the mean-field $\pi$F state there is a gauge flux of 
$\pi$ through every plaquette, and the mean-field Hamiltonian has the same form as 
Eq.~(\ref{eqn:mf-hamiltonian}) with $t = \Delta$. The discussion now proceeds almost 
identically to that for the staggered-flux state above. The universal physics of the $\pi$F state, 
including fluctuations, is encapsulated in the lattice gauge theory Hamiltonian
\begin{eqnarray}
\label{eqn:piF-hamiltonian}
&& {\cal H}_{\pi {\rm F}} = h \sum_{\langle \br \br' \rangle} e^2_{\br \br'}
- K \sum_{\square} \cos (\operatorname{curl} a) \\
&& - t \sum_{\br \in A} \, \sum_{\br' \rm{n. n.} \br}
\big[ (i  + (-1)^{(r_y - r'_y)}) f^\dagger_{\br \alpha} 
  e^{-i a_{\br \br'}} f^{\vphantom\dagger}_{\br' \alpha} + \text{H. c.} \big] \text{.} \nonumber
\end{eqnarray}
The low-energy effective field theory 
can be described by \emph{eight} massless two-component Dirac fermions minimally coupled to a 
noncompact $U(1)$ gauge field.  The field theory is thus the same as for the sF state, except 
that now $N_f= 8$.  

There are some differences between the $\pi$F and sF states.  First of all, the microscopic symmetries of the lattice model act rather differently on the continuum Dirac fields -- for the $\pi$F state, the symmetries are enumerated in Appendix A of Ref.~\onlinecite{stable-u1}.  Also, the $SU(2)$ spin version of the $\pi$F state is not a $U(1)$ spin liquid at all, but instead has a gapless $SU(2)$ gauge boson.\cite{xgw-qorder-ssl}  Finally, the model Eq.~(\ref{eqn:su4-hamiltonian}) has an additional discrete global ``charge-conjugation'' symmetry (called $C$) with no analog in $SU(2)$ spin models.  $C$ is defined as a particle-hole transformation of the spinon operators: $f^{\vphantom\dagger}_{\br \alpha} \to f^{\dagger}_{\br \alpha}$.  [Note that in an $SU(2)$ spin model this is equivalent to a particular $SU(2)$ spin rotation and is thus not a distinct symmetry.]

The $\pi$F state has an $SU(8)$ flavor symmetry, and
there is again a useful decomposition into the subgroup
\begin{equation}
SU(4)_{{\rm spin}} \times SU(2)_{{\rm nodal}} \subset SU(8)_{{\rm flavor}} \text{.}
\end{equation}
The spin $SU(4)$ is generated by the $T^a$, and the nodal $SU(2)$ is generated by the $\mu^i$ 
Pauli matrices.  We define the generators of $SU(8)$ to be ${\cal T}^A$, where $A = 1,\dots,63$.  
The ${\cal T}^A$ can be expressed as tensor products of the $T^a$ and $\mu^i$.  Proceeding as in the 
sF state, we define the 16-component fermion field $\Psi$, with all the flavor and Dirac indices suppressed.  [The notation here is identical to that of Ref.~\onlinecite{stable-u1}, except for the very minor difference that here the $SU(4)$ 
spin index is suppressed and $\Psi$ is written instead of $\Psi_\alpha$.]
The action of the $SU(8)$ symmetry is then $\Psi \to \exp(i \lambda^A {\cal T}^A) \Psi$.
As above, we can form the 64 bilinears with correlations enhanced by gauge fluctuations:
\begin{eqnarray}
N^A &=& -i \bar{\Psi} {\cal T}^A \Psi \text{,} \\
M &=& -i \bar{\Psi} \Psi \text{.}
\end{eqnarray}
As for the sF state, it it convenient to break the $N^A$ operators into three classes:
\begin{eqnarray}
N_A^{a,i} &=& -i \bar{\Psi} T^a \mu^i \Psi \text{,} \\
N_B^{a} &=& -i \bar{\Psi} T^a \Psi \text{,} \\
N_C^i &=& -i \bar{\Psi} \mu^i \Psi \text{.}
\end{eqnarray} 
The goal here is to find microscopic operators with the same transformation properties as the
continuum bilinears under the microscopic symmetries -- this is easily accomplished by making use of results in Ref.~\onlinecite{stable-u1}.  It should then be possible to 
numerically measure the correlations of these observables.

At this point, it would be natural to find spin operators built from $S^a_{\br}$ that 
transform as the various components of $N^A$ and $M$.  However, this is not the 
most convenient way to proceed, since the simulation of Ref.~\onlinecite{assaad} works 
directly in terms of the slave fermions and the microscopic gauge field used to 
decouple their quartic interaction.  If we restrict our attention to spin operators, 
most of the resulting observables are products of two or more $S^a_{\br}$ and are 
therefore of quartic or higher order in the fermions.  This is undesirable, most simply 
because such operators are rather difficult to deal with numerically.  Furthermore, 
it is conceivable (because $N_f=8$ may be rather large) that the microscopic slave 
fermions are good variables  and rather accurately represent the long-wavelength degrees of freedom.  
If this is the case, an operator quartic in $f_{\br \alpha}$ will have very little overlap 
with the continuum bilinears $N^A$ or $M$ -- precisely, the coefficient of these operators 
in Eq.~(\ref{eqn:eft-reln}) will be dominated by that of an appropriate four-fermion term.  
So while the bilinears should indeed give the dominant long-distance correlations, 
it may be necessary to go to unreasonably large distances to overcome the small prefactor.

\begin{table}
\begin{ruledtabular}
\begin{tabular}{c | c }
	$\text{  Field Theory  }$ & $\text{  Lattice  }$ \\
	\hline
	$N_A^{a,1}$ & $(-1)^{r_y} \big[ \exp\big((-1)^{r_x + r_y} \frac{3 i \pi}{4} \big) W_{\br, \br + \by}
		 f^\dagger_{\br \alpha} T^a_{\alpha \beta} f^{\vphantom\dagger}_{\br +\by, \beta}$ \\
	 & $+ \text{ H. c.} \big]$ \\
	 \hline
	 $N_A^{a,2}$ & $(-1)^{r_x} \big[ \exp\big( (-1)^{r_x + r_y} \frac{i \pi}{4} \big) W_{\br, \br + \bx}
	 	f^\dagger_{\br \alpha} T^a_{\alpha \beta} f^{\vphantom\dagger}_{\br + \bx, \beta}$ \\
	 & $+ \text{ H. c.} \big]$ \\
	 \hline
	 $N_A^{a, 3}$ & $(-1)^{(r_x + r_y)} f^\dagger_{\br \alpha} 
	 	T^a_{\alpha\beta} f^{\vphantom\dagger}_{\br \beta}$ \\
	 \hline
	 $N_B^a$ & $(-1)^{(r_x + r_y)} \big[ (W_{1,3} f^\dagger_{1 \alpha} T^a_{\alpha\beta} 
	 	f^{\vphantom\dagger}_{3 \beta} + \text{ H. c.})$ \\
	& $ \qquad\qquad\,\,\,\,\, +\, 
		(W_{2, 4} f^\dagger_{2\alpha} T^a_{\alpha\beta} f^{\vphantom\dagger}_{4 \beta} +
		\text{ H. c.}) \big]$ \\
	\hline
	$N_C^1$ & $(-1)^{r_y} \big[ \exp\big((-1)^{r_x + r_y} \frac{3 i \pi}{4} \big) W_{\br, \br + \by}
	 	f^\dagger_{\br \alpha} f^{\vphantom\dagger}_{\br +\by, \alpha}$ \\
	&	 $+ \text{ H. c.} \big]$ \\
	\hline
	$N_C^2$ & $(-1)^{r_x} \big[ \exp\big( (-1)^{r_x + r_y} \frac{i \pi}{4} \big) W_{\br, \br + \bx}
	 	f^\dagger_{\br \alpha} f^{\vphantom\dagger}_{\br + \bx, \alpha}$ \\
	& $+ \text{ H. c.} \big]$ \\
	\hline
	$N_C^3$ & $(-1)^{(r_x + r_y)} f^\dagger_{\br \alpha} f^{\vphantom\dagger}_{\br \alpha}$ \\
	\hline
	$M$ & $(-1)^{(r_x + r_y)} \big[ (W_{1,3} f^\dagger_{1 \alpha}  
	 	f^{\vphantom\dagger}_{3 \alpha} + \text{ H. c.})$ \\
	& $ \qquad\qquad\,\,\,\,\, +\, 
		(W_{2, 4} f^\dagger_{2\alpha}  f^{\vphantom\dagger}_{4 \alpha} +
		\text{ H. c.}) \big]$
\end{tabular}
\end{ruledtabular}
\caption{\label{tab:enhanced-su4-ops}Continuum fermion bilinears with enhanced correlations and their lattice counterparts for the $\pi$-flux state of the $SU(4)$ Heisenberg model discussed in the text.  The factors $W_{\br, \br'}$ encode the dependence on the lattice vector potential and are defined in the text.
As in Table~\ref{tab:enhanced-ops}, for $N^a_B$ and $M$ we consider the plaquette with lower-left corner at $\br$ and label the sites around it with the numbers $1,\dots,4$, proceeding counterclockwise from $\br$.}
\end{table}

To avoid these problems, we instead consider bilinears of the lattice slave fermions.\footnote{We
thank P. A. Lee for suggesting this to us.}  In most cases these involve products of fermions 
on different lattice sites, so in order to write gauge invariant operators it is necessary to 
include an appropriate dependence on the vector potential.  In Table~\ref{tab:enhanced-su4-ops} 
we enumerate lattice bilinears and their continuum counterparts -- in addition to transforming 
identically under the microscopic symmetries, the lattice observables also have the desirable 
property that they reduce exactly to the corresponding continuum operator upon taking the na\"{\i}ve continuum limit.  It is important to note that the form of these operators depends on the presence of the explicit background flux in the Hamiltonian, Eq.~(\ref{eqn:piF-hamiltonian}).  This background is not present in Ref.~\onlinecite{assaad}; this is simply due to a different choice of gauge field, which we denote $\tilde{a}_{\br \br'}$.  The two gauge fields are related by
\begin{eqnarray}
a_{\br, (\br + \bx)} &=& \tilde{a}_{\br, (\br + \bx)} + (-1)^{r_x + r_y} \Big(\frac{\pi}{4} \Big) \text{,}\nonumber \\
a_{\br, (\br + \by)} &=& \tilde{a}_{\br, (\br + \by)} + (-1)^{r_x + r_y} \Big(\frac{3 \pi}{4} \Big) \text{.}
\end{eqnarray}
In terms of $\tilde{a}_{\br \br'}$, the hopping term in the Hamiltonian Eq.~(\ref{eqn:piF-hamiltonian}) becomes
\begin{equation}
- \sqrt{2} t \sum_{\langle \br \br' \rangle} \big( f^{\dagger}_{\br \alpha} e^{-i \tilde{a}_{\br \br'}}
f^{\vphantom\dagger}_{\br' \alpha} + \text{H. c.} \big) \text{.}
\end{equation}

The factor $W_{\br, \br'}$ in Table~\ref{tab:enhanced-su4-ops} is a function of the vector 
potential included to keep the lattice bilinears gauge-invariant.  For $\br$ and $\br'$ nearest neighbors 
we simply have the usual exponential
\begin{equation}
W_{\br, \br'} = \exp(-i a_{\br, \br'}) \text{.}
\end{equation}
We also need to define $W$ for $\br$ and $\br'$ next-nearest neighbors.  In that case, to retain 
as much symmetry as possible we must sum over the two shortest paths connecting $\br$ and $\br'$.  
That is, if $\br' = \br + \bx + \by$, we define
\begin{eqnarray}
W_{\br, \br'} &=& \exp\big(-i a_{\br,(\br+\bx)} - i a_{(\br+\bx),(\br+\bx+\by)}\big) \\
&+& \exp\big(-i a_{\br,(\br+\by)} - i a_{(\br+\by),(\br+\bx+\by)}\big) \text{.} \nonumber
\end{eqnarray}

While it is not directly useful for obtaining numerical results, it is interesting to discuss the orders in the $\pi$F state in more physical terms.  It is easily seen that $N^{a,3}_A$ is the $SU(4)$ version of the N\'{e}el vector.  Also, $N^{1,2}_C$ together form the order parameter for the columnar and box VBS states, as in the sF case.  Specifically one finds
\begin{eqnarray}
(-1)^{r_y} S^a_{\br} S^a_{\br + \by} &\sim& N^1_C \text{,}\\
(-1)^{r_x} S^a_{\br} S^a_{\br + \bx} &\sim& N^2_C \text{.}
\end{eqnarray}
$N^3_C$ is an order parameter for a kind of $C$-breaking state that has been studied for a large class of $SU(N)$ magnets.\cite{C-breaking}  It should be noted that $N^3_C$ breaks time reversal as defined in Ref.~\onlinecite{stable-u1} (there time reversal sends $S^a \to - S^a$), but does not in the conventions of Ref.~\onlinecite{C-breaking}.  Finally, as in the sF case, $M$ transforms identically to a Chern-Simons term for the gauge field, and if spontaneously generated will lead to a chiral spin liquid.

\section{Consequences for Underdoped Cuprates}
\label{sec:cuprates}

In our view, one of the more promising routes toward a theoretical understanding 
of the underdoped cuprates views the pseudogap regime as a doped \emph{spin-liquid} Mott insulator.  
This line of thinking began with the ideas of Anderson\cite{pwa} and has subsequently been developed by many others.\cite{sachdev-review, lee-nagaosa-wen-review}

\begin{figure}
\includegraphics[width=6cm]{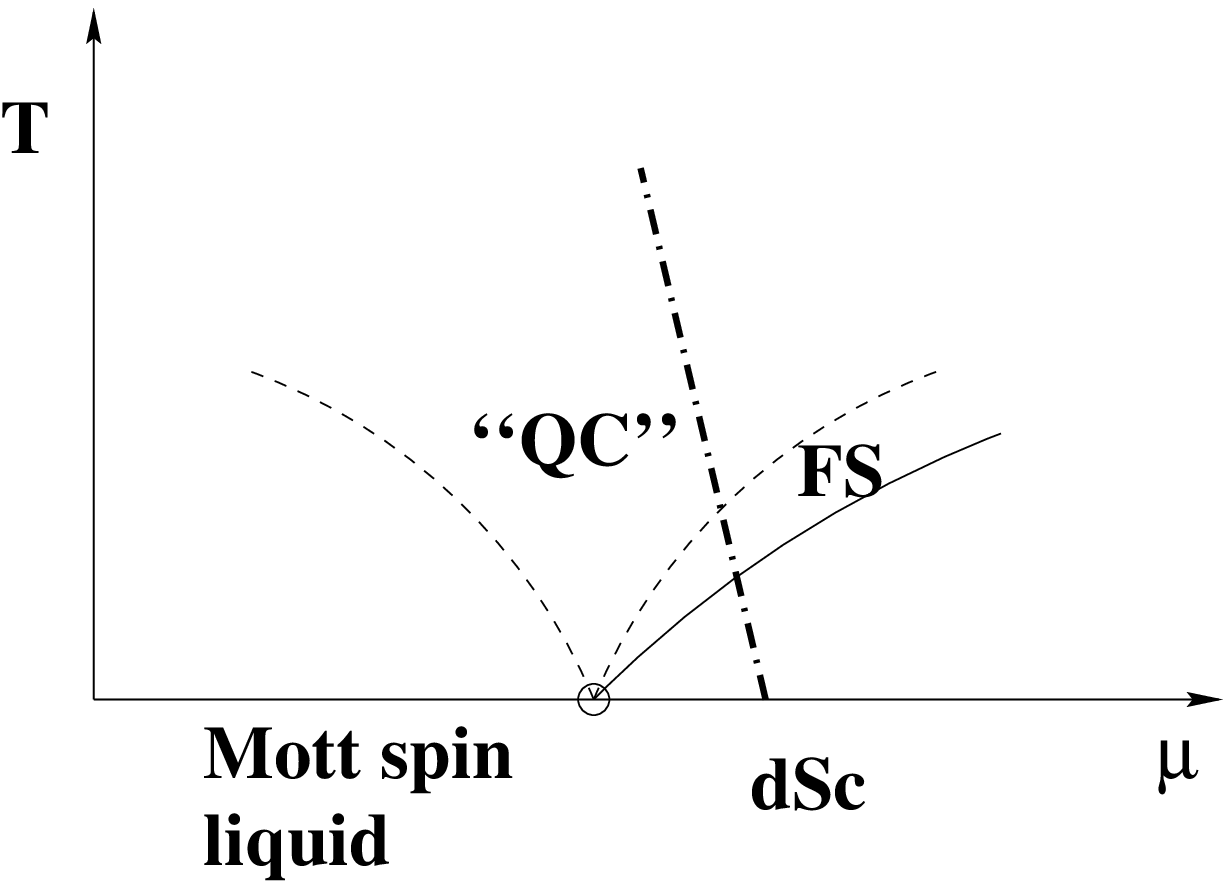}
\caption{Schematic temperature-chemical potential phase diagram of an underdoped cuprate superconductor.  For $\mu < \mu_c$ the holes are gapped and the ground state is a spin liquid Mott insulator, while when $\mu > \mu_c$ a finite density of holes has entered the system and the ground state is a $d$-wave superconductor.  The regime labeled ``QC'' is controlled by the Mott quantum critical point separating these two phases, while the FS regime is best thought of in terms of fluctuating $d$-wave superconductivity.  The dashed line represents the finite-temperature behavior of a real underdoped material (with a superconducting ground state) at fixed doping.}
\label{fig:phasediag}
\end{figure}

The current state of these ideas has been discussed recently by 
Senthil and Lee.\cite{senthil-lee}  For our present purposes, the key point of this 
picture is the close proximity of the underdoped $d$-wave superconductor to a Mott transition 
to a spin liquid insulator.  This is most clearly understood by thinking about the phase 
diagram as a function of hole chemical potential and temperature, as shown in 
Fig.~\ref{fig:phasediag} (from Ref.~\onlinecite{senthil-lee}).  Consider an underdoped material in the superconducting state, and imagine raising the temperature -- this is  represented by the dashed line in Fig.~\ref{fig:phasediag}.  Above $T_c$ there 
is a region that can be described as a phase-fluctuating  $d$-wave superconductor (labeled by ``FS'' in Fig.~\ref{fig:phasediag}), and at 
still higher temperatures the physics is controlled by the quantum critical point between the 
spin liquid and the superconductor.  We shall be interested in this ``high-temperature pseudogap'' region.

We take the Mott insulator in question to be the staggered flux 
algebraic spin liquid.\cite{wen-lee-su2,kim-lee,rantner-wen-prl,RWspin}  Furthermore, we consider the 
simplest scenario where this state is a stable phase.  In this case, the Mott insulating part 
of the phase diagram is controlled by a critical state with a dynamic critical exponent $z = 1$.  
The Mott transition is described by the condensation of a doubly-degenerate parabolic band of 
charge $e$ bosonic holons.  An important point is that the charge sector essentially has $z=2$, 
and at long distances the holon velocity goes to zero while the spinon and photon velocities go to a constant.  
This means that the holons move very slowly and couple only weakly to the spin sector without strongly 
influencing it.  Therefore the spin sector of the quantum critical regime should not change substantially 
from the finite-temperature physics of the algebraic spin liquid.  It is very important to note that this conclusion will hold only at temperatures above the crossover to the fluctuating superconductor (or ``FS'') regime.  Below this temperature, the coupling between the spin and charge sectors will be very important.  

It should then be possible to test whether the staggered flux algebraic spin liquid is relevant 
to the cuprates by probing the high-temperature pseudogap for signatures of the ASL fixed point.  
What are the upper and lower temperature scales defining this region?  Physically, one needs a 
large window of length scales where the spin sector is coherent and controlled by the ASL fixed point, 
but the charge sector is rather incoherent and sufficiently far from condensing that it does not 
substantially influence the spin sector.  Presumably the upper temperature is the pseudogap 
scale $T^*$, below which the Knight shift decreases, signaling the formation of spin singlets.  Because the uniform magnetization is a conserved density (one of the $J^a_\mu$), it can be seen by a standard scaling argument that the Knight shift in the ASL is expected to decrease \emph{linearly} with decreasing temperature.\footnote{This conclusion holds under the reasonable assumption that there is not some other, more relevant operator that is also symmetry-equivalent to the $\bq = 0$ spin density.} 
The lower temperature scale is presumably bounded below by $T_\nu$, the onset temperature for the Nernst signal.\cite{ong-nernst-nature, ong-nernst-prb}  
The onset of the Nernst signal can probably be identified with the onset of a substantial degree of 
phase coherence in the charge sector.

To take a specific example, in YBa$_2$Cu$_3$O$_{6 + x}$
$T_{\nu}$ is about 110 K for $x = 0.5$,\cite{wang-thesis} while, for an $x = 0.53$ sample, the Knight shift increases linearly from this temperature up to 300 K (the highest temperature measured).\cite{alloul-nmr}   So in this material it is reasonable to look for algebraic spin liquid physics at least in the range
$110 {\rm K} < T < 300 {\rm K}$, and perhaps at even higher temperature.  This window is expected to grow with underdoping, which increases $T^*$ and decreases $T_\nu$.\cite{ong-nernst-prb}

Within this temperature range the properties of the algebraic spin liquid can be probed by looking 
at the dynamic correlations for frequencies $\omega \ll c J$.  The energy $c J$, where $J$ is the 
exchange energy and $c$ is a number of order unity, plays the role of a high-energy cutoff above 
which the physics is presumably nonuniversal.
If one further restricts to frequencies $\omega \gg T$, it is possible to probe the zero-temperature 
critical ground state and avoid complicated issues of critical dynamics.  Furthermore, 
various quantities will exhibit critical scaling for all frequencies $\omega \ll c J$.  
It is likely that the simplest test of this physics would be to look for scaling in 
the $\bQ = (\pi,\pi)$ magnetic neutron scattering. Specifically the 
imaginary part of the dynamic spin susceptibility at $(\pi, \pi)$ is expected to 
satisfy
\begin{equation}
\chi''(\bq, \omega, T) \sim \frac{1}{\big((\bq - \bQ)^2 + \omega^{2}\big)^{\frac{2- \eta}{2}}}
f\left(\frac{|\bq - \bQ|}{T}, \frac{\omega}{T}\right).
\end{equation} 
The exponent $\eta$ is not known. A rough guess may be obtained from studies of 
variational wavefunctions for the sF spin liquid. 
In particular the well-studied projected nearest-neighbor $d$-wave BCS state might be expected to 
capture the physics of the sF spin liquid. From the known result\cite{arun-wavefn, ivanov-thesis} on the 
equal-time spin correlations in that 
wavefunction one extracts $\eta \approx 0.5$. 
The scaling form above also has direct implications for the NMR relaxation rate $1/T_1$ at the Cu site 
[which is sensitive to the $(\pi, \pi)$ spin correlations]. We have $1/T_1 \sim T^\eta$ with 
$\eta$ roughly about 0.5.
 
If this type of scaling is seen, it would be important to think about whether the $SU(4)$ symmetry 
can be explicitly tested by probing the other observables with enhanced correlations. In particular 
the power law VBS correlations can possibly be looked for. 

Very recently, scaling in the $\bQ = (\pi,\pi)$ magnetic neutron scattering has been observed in underdoped samples of  YBa$_2$Cu$_3$O$_{6 + x}$, for both $x=0.5$ ($T_c = 59 K$) and $x=0.35$ ($T_c = 18 K$).\cite{buyers-kitp-seminar}  In both samples $\omega / T$ scaling is seen in $\chi''(\bq,\omega,T) / \chi''(\bq, \omega, T=0)$.  Further analysis of the data is needed to understand whether this scaling may be related to algebraic spin liquid physics.  

We note that scaling has also been reported in very lightly doped 
La$_{2-x}$Sr$_x$CuO$_4$ (Refs.~\onlinecite{keimer-prl} and~\onlinecite{keimer-prb})  and more recently in Li-doped La$_2$CuO$_4$ (Ref.~\onlinecite{bao}). Specifically the $\bq$-integrated scattering intensity (which is dominated by the signal near $\bQ$) shows $\omega/T$  scaling. However, the prefactor decreases with increasing frequency unlike that expected from the scaling 
form above.\footnote{We note that conformal invariance and unitarity require $\eta > 0$.  See Ref.~\onlinecite{cft-review} and references therein.} Further, there is no sign of scaling in 
the momentum dependence of the scattering. We therefore think it unlikely that the $\omega/T$ scaling reported 
in Refs.~\onlinecite{keimer-prl,keimer-prb,bao} is due to any underlying algebraic spin liquid. However, it would be very interesting to look for scaling in moderately doped samples in the high-temperature pseudogap regime (as opposed to the very 
lightly doped samples studied in Refs.~\onlinecite{keimer-prl} and~\onlinecite{keimer-prb}). 

It is important to note that, even in this relatively simple picture where 
the algebraic spin liquid is stable, the critical scaling may be modified by the 
presence of weakly irrelevant perturbations.  Particularly worrisome is the fermion 
velocity anisotropy, which is known to be rather large in the superconductor.  However, 
it is not at all clear that the low-temperature anisotropy in the superconductor is to be 
identified with the anisotropy in the spin sector at high temperatures.  It may well be the 
case that the spin sector flows close to the isotropic ASL fixed point, but at lower temperatures 
the charge sector could induce the anisotropy that obtains in the ground state.  Similar issues arise regarding the location of the nodal points in momentum space -- in the sF state these are fixed at $\bq = (\pi/2, \pi/2)$, but this is not the case in the superconducting state. While these issues 
remain somewhat mysterious at present, we would like to emphasize that the whole approach of thinking 
about the underdoped cuprates in terms of an algebraic spin liquid is only useful to the extent 
that one comes near the fixed point, which is isotropic.

\section{Discussion}
\label{sec:discussion}

In the analysis of this paper, we have assumed that the most relevant operators in the sF and $\pi$F spin liquids are the $N^a$ and $M$ fermion bilinears;  these operators therefore give rise to the dominant power-law correlations.  While this scenario will certainly hold for sufficiently large $N_f$, it is not known whether it continues to hold for the interesting cases of $N_f = 4$ or $8$.  In particular, monopole operators carry nontrivial flavor (and other) quantum numbers,\cite{bkw, monopole-qn} and if some such multiplet of operators becomes more relevant than, say, $N^a$, it will dominate the long-distance correlations.  We note that if this does in fact happen, the main features of the results discussed here will still hold.  Specifically, there will be a set of superficially unrelated and slowly-varying competing orders all exhibiting the same power-law decay and unified the by the emergent $SU(N_f)$ symmetry.  The observables involved will, however, be different from those discussed here.  Furthermore, even if $N^a$ and $M$ do not give rise to the \emph{dominant} competing orders, they will still give rise to slowly-decaying correlations.  It should be noted that these issues do not affect the prediction of $\omega/T$ scaling in the $\bq = (\pi,\pi)$ magnetic scattering.

Many aspects of the staggered flux spin liquid state are
very reminiscent of the physics of the more familiar one-dimensional critical spin liquids. 
The most striking similarity  
is perhaps in the criticality itself; indeed, the staggered flux and other algebraic spin liquids can be stable critical phases in two dimensions,
much like their better-known one-dimensional counterparts. In both cases the spin correlations are described 
by nontrivial power laws with large anomalous dimensions. More technically, as discussed in detail in this paper, the sF spin liquid is conveniently analyzed in terms of fermionic $S = 1/2$ spinon variables. Similarly, a fermionic description is often 
a useful technical device in analyzing the physics of one-dimensional spin liquids. Finally, there is similarity in how the
semiclassical instantons are represented in terms of the fermions.
In $d = 1$, for instance in the antiferromagnetic XXZ $S = 1/2$ model, the semiclassical instantons are
just $2\pi$ phase slips. In a fermionic
representation (obtained via Jordan-Wigner transformation) 
these are umklapp processes where a right mover becomes a
left mover or vice versa.
In the $d =2$ spin model, the semiclassical instantons 
change the Skyrmion number associated with the N\'{e}el vector configuration.  
In the staggered flux spin liquid these are
again described as operators that move a fermionic spinon from one node to
the other (the northeast movers to the northwest movers, for instance). Perhaps these similarities can be exploited toward 
deepening our understanding of such nontrivial two-dimensional algebraic spin liquids. 

We note that the spin-charge-separated variables used to describe the algebraic spin liquids considered here have no obvious \emph{a priori} connection to the pattern of competing orders arising within these states, and it is remarkable 
that they lead to this kind of physics.  In fact, the observables with slowly-varying fluctuations 
correspond to \emph{bilinears} of the fermions.  It is not known how to formulate a field theory for 
the sF state where these variables are in some sense the fundamental fields.  This points out that, 
in doing phenomenological modeling of strongly correlated systems, one should be cautious about 
simply introducing new fields by hand for the slowly fluctuating observables -- these are not 
necessarily the variables that will naturally lead to a correct description of the underlying physics.

Many theoretical issues remain to be addressed if a solid connection is to be made between the picture advocated here and experiments in the cuprates.  We feel the most serious of these involve coupling to the charge sector, which so far has not been carefully taken into account.  Specifically, it will be important to understand the physics of the doped algebraic spin liquid at all temperatures, not only in the range where the spin sector should be controlled by the undoped fixed point.  The zero-temperature fate of a doped ASL is a question that also merits further exploration.  The most common view has been that $d$-wave superconductivity obtains immediately at $T=0$ upon introduction of a finite density of charge carriers.  It is particularly intriguing to ask whether some of the exotic character of the ASL can survive down to $T=0$ even in the presence of doped holes, possibly leading to exotic \emph{metallic} states.

\acknowledgments{The authors gratefully acknowledge discussions with F. F. Assaad, L. Balents, W. J. L. Buyers, P. A. Lee, J. B. Marston, S. Sachdev, and X.-G. Wen.  This research is supported by the Department of Defense NDSEG program (M. H.), NSF Grant No. DMR-0308945 (T. S.), and NSF Grant Nos. DMR-0210790 and PHY-9907949 (M. P. A. F.).  T. S. also acknowledges funding from the NEC Corporation, the Alfred P. Sloan Foundation, and an award from the The Research Corporation.
M. H. is grateful for the hospitality of the Aspen Center for Physics and the MIT condensed matter 
theory group, where some of this work was carried out.}

\appendix
\section{Continuum Fields and Microscopic Symmetries}
\label{app:cont-limit}

Here we provide a discussion of the continuum limit of the staggered-flux mean-field state, and the resulting action of the microscopic symmetries on the continuum fields.  The same procedure was discussed in Appendix A of Ref.~\onlinecite{stable-u1} for the $SU(N)$ $\pi$-flux state.  Much of the analysis is identical, but for completeness we reproduce it here.  It is important to note that the final results differ because the symmetries in the $\pi$F and sF states act differently on the lattice spinons and hence also on the continuum fields.

The starting point is the mean-field Hamiltonian of Eq.~(\ref{eqn:mf-hamiltonian}).  We choose the four-site unit cell labeled by $(\boldsymbol{R},i)$, with $\bR = 2 n_x \bx + 2 n_y \by$ and
$\br (\bR, i) = \bR + \boldsymbol{v}_i$, where
\begin{equation} 
\boldsymbol{v}_i = \left\{ \begin{array}{ll}
{\bf 0} \text{,} & i = 1\text{,} \\
\bx \text{,} & i = 2 \text{,} \\
\bx + \by \text{,} & i =3 \text{,} \\
\by \text{,} & i = 4 \text{.}
\end{array} \right.
\end{equation}
The spinon operator at site $(\bR, i)$ is denoted $f_{\bR i \alpha}$.

It is a trivial exercise to go to momentum space and solve Eq.~(\ref{eqn:mf-hamiltonian}); 
in the reduced Brillouin zone $k_x,k_y \in [0,\pi)$
one finds gapless Fermi points at $\boldsymbol{Q}_0 \equiv (\pi/2,\pi/2)$.  Near this point the dispersion can be described by 4 two-component Dirac fermions.  It is convenient to denote these by $\psi^A_{a \alpha}(\bR)$.  Here $a = 1,2$ and $\alpha = 1,2$ are the $SU(4)$ flavor indices [$\alpha$ is simply the $SU(2)$ spin index].  Also, $A =1,2$ labels the two components of each
spinor (this is usually suppressed).  These fields are related to the lattice spinons as follows:
\begin{eqnarray}
\psi^1_{1\alpha}(\bR) &\sim&
	\frac{1}{2\sqrt{2}\ell} e^{i\bQ_0\cdot \bR} (f_{\bR 1\alpha} + f_{ \bR 3\alpha}) \text{,} \\
\psi^2_{1\alpha}(\bR) &\sim&
	\frac{-i}{2\sqrt{2}\ell} e^{i \bQ_0 \cdot \bR} (f_{\bR 2\alpha} - f_{\bR 4\alpha}) \text{,} \\
\psi^1_{2\alpha}(\bR) &\sim&
	\frac{-e^{-i\pi/4}}{2\sqrt{2}\ell} e^{i\bQ_0 \cdot{\bf R}} (f_{\bR 2\alpha} + f_{\bR 4\alpha}) \text{,} \\
\psi^2_{2\alpha}(\bR) &\sim&
	\frac{-e^{-i\pi/4}}{2\sqrt{2}\ell} e^{i\bQ_0 \cdot \bR} (f_{\bR 1\alpha} - f_{\bR 3\alpha})\text{,}
\end{eqnarray}
where $\ell$ is the lattice spacing.

We can now set $t = \Delta$ to remove the velocity anisotropy, and put it back in as a perturbation as discussed in Sec.~\ref{sec:stability}.  Note that the above results do not depend on $t/\Delta$ and are identical to the case of the $\pi$-flux state.
In momentum space the continuum Hamiltonian takes the form
\begin{equation}
{\cal H}_c = \int \frac{d^2 q}{(2\pi^2)} \psi^\dagger_{a\alpha}({\bf q})
\big(q_1 \tau^1 + q_2 \tau^2\big) \psi^{\vphantom\dagger}_{a \alpha}({\bf q}) \text{,}
\end{equation}
where we have chosen units to set the velocity to unity, and $\tau^i$ are the usual Pauli matrices acting
in the two-component Dirac ``spin'' space.  Here we use the following rotated coordinates:
\begin{eqnarray}
\label{rotated-coords}
q_1 &=& \frac{1}{\sqrt{2}}(q_x + q_y) \text{,} \\
q_2 &=& \frac{1}{\sqrt{2}}(-q_x + q_y) \text{.}\nonumber
\end{eqnarray}

It is convenient to work with the eight-component object
\begin{equation}
\label{eqn:eight-comp}
\Psi = \left( \begin{array}{c}
\psi_{11} \\
\psi_{12} \\
\psi_{21} \\
\psi_{22}
\end{array} \right) \text{.}
\end{equation}
The generators of flavor $SU(4)$ can be expressed as tensor products of $SU(2)_{{\rm spin}}$ and $SU(2)_{{\rm nodal}}$ generators (see Sec.~\ref{sec:symmetries}).  Using the convention specified in Eq.~(\ref{eqn:eight-comp}) this can be expressed in matrix notation; for example,
\begin{equation}
\sigma^i \mu^j = \left( \begin{array}{cc}
(\mu^j)_{1 1} \sigma^i & (\mu^j)_{1 2} \sigma^i \\
(\mu^j)_{2 1} \sigma^i & (\mu^j)_{2 2} \sigma^i
\end{array} \right) \text{.}
\end{equation}

We now quote the action of the microscopic symmetries on the lattice and continuum fields, including the spinons as well as the magnetic flux and electric field.  To simplify the form of the results, we often make an additional global gauge transformation $f_{\br \alpha} \to e^{i \phi} f_{\br \alpha}$ in going from the lattice to the continuum transformation laws.  The lattice symmetries discussed below generate the full space group of the square lattice.

\emph{$x$-translations}.  Translations by one site in the $x$ direction act on the lattice spinons as follows:
\begin{eqnarray}
f_{\br \alpha} &\to& \epsilon_{\br} (i \sigma^2)_{\alpha\beta} f^\dagger_{\br+\bx,\beta} \text{,} \\
f^\dagger_{\br \alpha} &\to& \epsilon_{\br} (i \sigma^2)_{\alpha\beta} f_{\br+\bx,\beta}
\nonumber\text{,}
\end{eqnarray}
where
\begin{equation}
\epsilon_{\br} = \left\{ \begin{array}{ll}
	+1 & {\br} \in A \text{,} \\
	-1 & {\br} \in B \text{.}
\end{array} \right.
\end{equation}
The resulting continuum transformation law is
\begin{eqnarray}
\Psi &\to& \big[ \Psi^\dagger (i \tau^1) (i \sigma^2) \big]^T \text{,} \\
\Psi^\dagger &\to& \big[(i \sigma^2) (i \tau^1) \Psi \big]^T \text{.} \nonumber
\end{eqnarray}
The electric field and magnetic flux both change sign under translation by one lattice site.

\emph{Rotations}.  We choose to make a $\pi/2$ counterclockwise rotation 
about the point
$(\bx + \by)/2$, which lies at a plaquette center.  Under this operation we have
$\br \to \br' = (-r_y +1,r_x)$, and the action on the spinons is
\begin{equation}
f_{\br \alpha} \to \epsilon_{\br} f_{\br' \alpha} \text{.}
\end{equation}
In the continuum
\begin{equation}
\Psi(\bR) \to \exp\Big(\frac{i\pi}{2}\Big(\frac{\mu_1 + \mu_2}{\sqrt{2}}\Big)\Big)
	\exp\Big(\frac{i\pi}{4}\tau^3\Big) \Psi(\bR')\text{.}
\end{equation}
Under this operation, the electric field is a vector, and the magnetic flux is a scalar.

\emph{Reflections}.  We consider the reflection $\br \to \br' = (-r_x, r_y)$.  The spinons transform trivially:
\begin{equation}
f_{\br \alpha} \to f_{\br' \alpha} \text{,}
\end{equation}
resulting in the continuum expression
\begin{equation}
\Psi(\bR) \to (i \mu^2)\exp\Big(\frac{i\pi}{2}\Big(\frac{\tau^1+\tau^2}{\sqrt{2}}\Big)\Big) 
	\Psi(\bR') \text{.}
\end{equation}
Note that in rotated coordinates $\bR = (R_1, R_2) \to \bR' = (R_2, R_1)$.  The electric field transforms as a vector under reflections, and the magnetic flux as a pseudoscalar.

\emph{Time reversal}.  Time-reversal symmetry is implemented by the \emph{antiunitary} operation
\begin{eqnarray}
f_{\br \alpha} &\to& \epsilon_{\br} f^\dagger_{\br \alpha} \text{,} \\
f^\dagger_{{\br}\alpha} &\to& \epsilon_{\br} f_{\br \alpha} \nonumber \text{.}
\end{eqnarray}
The resulting continuum operation is
\begin{eqnarray}
\Psi &\to&  \big[ \Psi^\dagger (i \tau^3) (i \mu^3) \big]^T  \text{,} \\
\Psi^\dagger &\to& \big[ (i\tau^3)(i\mu^3) \Psi \big]^T \nonumber \text{.}
\end{eqnarray}
The electric field is \emph{odd} under time reversal, while the magnetic flux does not transform.  This reverses the more familiar situation of real electromagnetism, where electric charge is invariant under time reversal but magnetic charge is odd.

\section{large-$N_f$ Diagrammatics and RG}
\label{app:largen-diagrams}

The starting point for the large-$N_f$ expansion is simply na\"{\i}ve perturbation theory in the gauge interaction vertex.  The fermion propagator is
\begin{equation}
\feyn{fa\vertexlabel_{k}f} = \frac{1}{\kslash} = \frac{\kslash}{k^2} \text{,}
\end{equation}
where we have introduced the notation $\kslash = k_\mu \gamma_\mu$.
The bare photon propagator takes the form
\begin{equation}
\annotate{1.0}{-0.2}{\text{\large{O}}}
\feyn{\vertexlabel_\mu g\vertexlabel_{q}g\vertexlabel_{\nu}} =  \frac{\delta_{\mu \nu} + (\frac{16 \xi |q|}{N_f e^2} - 1) \frac{q_\mu q_\nu}{q^2}}{(q^2/e^2)} \text{.}
\end{equation}
The rather unusual momentum dependence of the numerator is due to our choice of a ``nonlocal'' gauge -- this choice is made purely for technical convenience as it results in a simpler form for the photon propagator at leading order in $1/N_f$.  Finally we have the vertex
\begin{equation}
\Diagram{& fu \\ \vertexlabel_{\mu}g \\ & fd}
\annotate{-0.35}{0.25}{\blacktriangledown}
\annotate{-0.35}{-0.44}{\blacktriangledown}
 = - \gamma_\mu
\end{equation}

The next step in constructing the large-$N_f$ perturbation theory is to calculate the leading order photon propagator.  Recalling that $e^2 \sim 1/N_f$, 
it is easy to see that the leading contribution represented by the geometric series
\begin{eqnarray}
\feyn{gg} &=& \bpprop + \bpprop\fermiloop\bpprop \\
&+& \bpprop\fermiloop\bpprop\fermiloop\bpprop + \cdots \nonumber \text{.}
\end{eqnarray}
Upon summing the series and taking the limit of small $q$, the full photon propagator is
\begin{equation}
\feyn{\vertexlabel_{\mu}g\vertexlabel_{q}g\vertexlabel_{\nu}}  = \frac{16}{N_f |q|} \Big( \delta_{\mu \nu} + (\xi - 1) \frac{q_\mu q_\nu}{q^2} \Big) +
{\cal O}(1/N_f^2)
\label{eqn:full-photon-prop}
\end{equation}

The perturbation series for any desired correlator is then easily built out of the $1/N_f$ photon propagator Eq.~(\ref{eqn:full-photon-prop}), the fermion propagator, and the vertex.  For example, the fermion Green's function $\langle \Psi(k) \bar{\Psi}(k') \rangle = (2 \pi)^3 \delta(k - k') G^{(2)}(k)$ is represented as
\begin{equation}
\fermiprop + \fermiprop\feyn{fglf}\annotate{-1.0}{-0.1}{\blacktriangleright} \fermiprop + {\cal O}(1/N_f^2)
\text{.}
\end{equation}

As discussed in Sec.~\ref{sec:largen}, we can implement a renormalization group using the large-$N_f$ expansion.  As an example, consider the fermion Green's function and suppose we have added a single perturbation to the Lagrangian, represented by the coupling $g$.  It is most convenient to explicitly keep track only of the anomalous part of the scaling -- that is, we implicitly subtract off all contributions to the Callan-Symanzik equation that give rise to the engineering dimensions of fields and coupling constants.  This is denoted by writing primed versions of the appropriate quantities; for example, the engineering dimension of $\Psi$ is unity, so, denoting the scaling dimension of $\Psi$ by $\Delta_\Psi$, we write $\Delta_\Psi = 1 + \Delta'_\Psi$.  The Callan-Symanzik equation then takes the form
\begin{equation}
\Big[ - \Big(\frac{\partial}{\partial\ell}\Big)' + 2 \Delta'_\Psi + \Big( \frac{\partial g}{\partial \ell} \Big)' \frac{\partial}{\partial g} \Big]
G^{(2)}(k) = 0 \text{.}
\end{equation}

\section{Irrelevance of the Velocity Anisotropy}
\label{app:anisotropy}

In this appendix we show that the velocity anisotropy for the fermions is irrelevant at the algebraic spin liquid fixed point, at least to leading order in $1/N_f$.  We do this by calculating the coefficient of $\delta / N_f$ in $d \delta / d \ell$.  In Refs.~\onlinecite{vafek-anisotropy} and~\onlinecite{ftv-longpaper}, the same RG flow was calculated to leading order in $1/N_f$ but for arbitrary $\delta$.  Our calculation is essentially equivalent to that of Refs.~\onlinecite{vafek-anisotropy} and~\onlinecite{ftv-longpaper}; the only difference is that we are interested here only in local stability and can work perturbatively in $\delta$ from the beginning, which simplifies some technical aspects and makes the presentation more striaightforward.  Our results are in complete agreement with those of Refs.~\onlinecite{vafek-anisotropy} and~\onlinecite{ftv-longpaper}.

We employ the renormalization group approach discussed in Sec.~\ref{sec:largen}.  We need to calculate the Green's function $G^{(2)}(k)$, keeping terms of order $\delta$, $1/N_f$, and $\delta/N_f$, and then apply the appropriate Callan-Symanzik equation to determine $d \delta / d \ell$.  For ease of presentation we work in Feynman gauge ($\xi = 1$).  We have also carried out these calculations in an arbitrary covariant gauge with no effect on the final result.

The anisotropy was discussed in Sec.~\ref{sec:stability}, and the perturbation to the Lagrangian can be written
\begin{equation}
K_a = -i \delta \bar{\Psi} \mu^3 \hat{\gamma}_\mu (\partial_\mu + i a_\mu) \Psi \text{,}
\end{equation}
where we have introduced the notation $\hat{\gamma}_\mu = \gamma_1 \delta_{\mu,1} - \gamma_2 \delta_{\mu, 2}$.
This term is represented by two vertices.  The first is the correction to the fermion kinetic energy:
\begin{equation}
\feyn{ff}\annotate{-1.5}{-0.1}{\blacktriangleright}\annotate{-0.5}{-0.1}{\blacktriangleright}\annotate{-1.0}{-0.1}{\boxtimes} = -\delta  \mu^3 (k_1 \gamma_1 - k_2 \gamma_2) = - \delta \mu^3 \hat{\kslash} \text{.}
\end{equation}
Here $\hat{\kslash} \equiv (k_1 \gamma_1 - k_2 \gamma_2)$.  The second is the correction to the vertex:
\begin{equation}
\Diagram{& fu \\ \vertexlabel_{\mu}g \\ & fd}
\annotate{-0.35}{0.25}{\blacktriangledown}
\annotate{-0.35}{-0.44}{\blacktriangledown}
\annotate{-0.68}{-0.15}{\boxtimes}
 = - \delta \mu^3 \hat{\gamma}_\mu \text{.}
\end{equation}

We need to calculate the order $1/N_f$ contributions to the fermion self-energy, up through linear order in $\delta$.  These are given by a sum of four diagrams
\begin{equation}
\Sigma^{(1/N_f)}(k) = \sum_{i = 0}^3 \Sigma_i (k) \text{.}
\end{equation}
Here $\Sigma_0$ is the isotropic (\emph{i.e.} $\delta = 0$) contribution
\begin{equation}
\Sigma_0(k) = \feyn{f\vertexlabel_{k+q}glf}\annotate{-1.0}{1.2}{q}\annotate{-1.0}{-0.1}{\blacktriangleright}
= \frac{16}{N_f} \int \frac{d^3 q}{(2\pi)^3} \frac{1}{|q|} \Big[ \gamma_\mu \frac{1}{(\kslash + \qslash)} \gamma_\mu \Big] 
\label{eqn:sigma0}
\end{equation}
The other three diagrams are the anisotropy contributions.  We suppress the momentum labels since the momentum structure is the same as Eq.~(\ref{eqn:sigma0}).  We have
\begin{eqnarray}
\Sigma_1(k) &=& \feyn{fglf}
\annotate{-1.5}{-0.1}{\blacktriangleright}
\annotate{-0.5}{-0.1}{\blacktriangleright}\annotate{-1.0}{-0.1}{\boxtimes} \\
&=& - \frac{16 \delta \mu^3}{N_f} \int \frac{d^3 q}{(2\pi)^3} \frac{1}{|q|}
\Big[ \gamma_\mu \frac{1}{(\kslash + \qslash)} (\hat{\kslash} + \hat{\qslash}) \frac{1}{(\kslash + \qslash)}
\gamma_\mu \Big] \nonumber \\
\Sigma_2(k) &=& \feyn{fglf}
\annotate{-1.0}{-0.1}{\blacktriangleright}
\annotate{-2.0}{-0.1}{\boxtimes} \\
&=& \frac{16 \delta \mu^3}{N_f} \int \frac{d^3 q}{(2 \pi)^3} \frac{1}{|q|}
\Big[ \gamma_\mu \frac{1}{(\kslash + \qslash)} \hat{\gamma}_\mu \Big] \nonumber \\
\Sigma_3(k) &=& \feyn{fglf}
\annotate{-1.0}{-0.1}{\blacktriangleright}
\annotate{0}{-0.1}{\boxtimes} \\
&=& \frac{16 \delta \mu^3}{N_f} \int \frac{d^3 q}{(2 \pi)^3} \frac{1}{|q|}
\Big[ \hat{\gamma}_\mu \frac{1}{(\kslash + \qslash)} \gamma_\mu \Big] \nonumber
\end{eqnarray}

We evaluate these integrals using dimensional regularization, which introduces a mass scale $\mu$ that roughly plays the role of a UV cutoff.  Keeping track only of the logarithmically divergent parts, the results are
\begin{equation}
\Sigma_0(k) = \frac{8}{3 \pi^2 N_f} \kslash \operatorname{ln} (|k|/\mu)
\end{equation}
and
\begin{equation}
\Sigma_1(k) + \Sigma_2(k) + \Sigma_3(k) = - \frac{152 \, \delta \mu^3}{15 \pi^2 N_f} \hat{\kslash} 
\operatorname{ln} (|k|/\mu) \text{.}
\end{equation}

We can now calculate $G^{(2)}(k)$ to the appropriate order and use the following Callan-Symanzik equation to determine the flow of the anistropy:
\begin{equation}
\Big[ - \Big(\frac{d}{d \ell}\Big)' + 2 \Delta'_\Psi + \Big( \frac{d \delta}{d \ell} \Big)' \frac{\partial}{\partial \delta} \Big] G^{(2)}(k) = 0
\label{eqn:delta-cs}
\end{equation}
The diagrams contributing to the Green's function are
\begin{eqnarray}
G^{(2)}(k) &=& \fermiprop +
\feyn{ff}
\annotate{-1.5}{-0.1}{\blacktriangleright}
\annotate{-0.5}{-0.1}{\blacktriangleright}
\annotate{-1.0}{-0.1}{\boxtimes}
+
\fermiprop\feyn{fglf}\annotate{-1.0}{-0.1}{\blacktriangleright}\fermiprop \nonumber  \\
&+& 
\feyn{ff}
\annotate{-1.5}{-0.1}{\blacktriangleright}
\annotate{-0.5}{-0.1}{\blacktriangleright}
\annotate{-1.0}{-0.1}{\boxtimes}
\feyn{fglf}\annotate{-1.0}{-0.1}{\blacktriangleright}\fermiprop \nonumber \\
&+& \fermiprop\feyn{fglf}\annotate{-1.0}{-0.1}{\blacktriangleright}
\feyn{ff}
\annotate{-1.5}{-0.1}{\blacktriangleright}
\annotate{-0.5}{-0.1}{\blacktriangleright}
\annotate{-1.0}{-0.1}{\boxtimes} \nonumber \\
&+& \fermiprop\feyn{fglf}
\annotate{-1.5}{-0.1}{\blacktriangleright}
\annotate{-0.5}{-0.1}{\blacktriangleright}\annotate{-1.0}{-0.1}{\boxtimes}\fermiprop \nonumber \\
&+& \fermiprop\feyn{fglf}
\annotate{-1.0}{-0.1}{\blacktriangleright}
\annotate{-2.0}{-0.1}{\boxtimes}\fermiprop \nonumber \\
&+& \fermiprop\feyn{fglf}
\annotate{-1.0}{-0.1}{\blacktriangleright}
\annotate{0}{-0.1}{\boxtimes}\fermiprop \text{,}
\end{eqnarray}
and therefore
\begin{eqnarray}
G^{(2)}(k) &=& \frac{1}{\kslash} \Big[ 1 + \frac{8}{3 \pi^2 N_f} \operatorname{ln} (|k|/\mu) \Big] \\
&-& \delta \mu^3 \frac{1}{\kslash} \hat{\kslash} \frac{1}{\kslash}
\Big[ 1 + \frac{232}{15 \pi^2 N_f} \operatorname{ln} (|k|/\mu) \Big] \nonumber \text{.}
\end{eqnarray}

Noting that $(d \delta / d \ell)' = (d \delta / d \ell)$ because $\delta$ is marginal at infinite $N_f$, it is  straightforward to apply Eq.~(\ref{eqn:delta-cs}) to obtain $\Delta'_\Psi = -4 / 3 \pi^2 N_f$ and
\begin{equation}
\frac{d \delta}{d \ell} = - \frac{64}{5 \pi^2 N_f} \delta \text{.}
\end{equation}

\section{Two-point correlations of $N^a$ and $M$ to all orders in $1/N$}
\label{app:allorders}

\begin{figure}
\includegraphics[width=8cm]{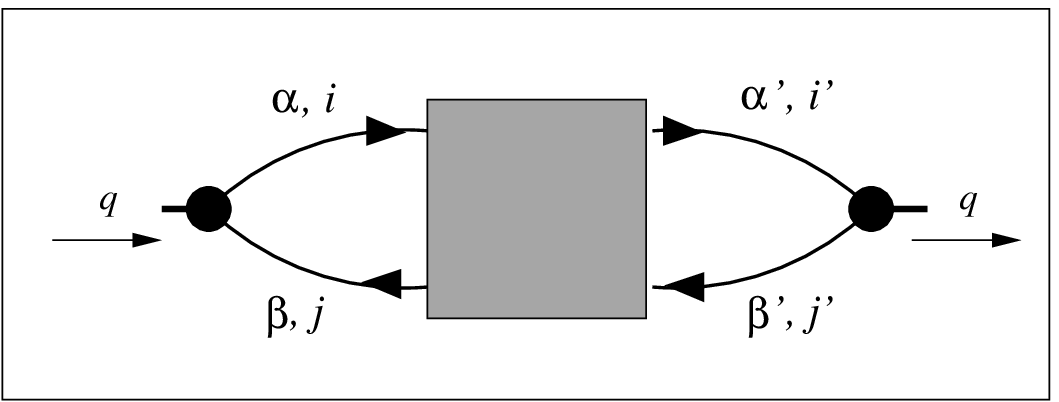}
\caption{Representation of an arbitrary diagram contributing to the two-point function of a fermion bilinear.  The two dark circles represent the bilinear's matrix structure and are positioned at the points where the bilinear is inserted into the diagram.  The shaded region is built from the ingredients of the large-$N_f$ perturbation theory described in Sec.~\ref{sec:largen} -- fermion and photon lines, and the gauge vertex.  The pairs $(\alpha, i)$ denote the $SU(N_f)$ flavor ($\alpha = 1,\dots,N_f$) and Dirac spin ($i = 1,2$) indices of the fermions.}
\label{fig:bilinear-2pt}
\end{figure}

In this appendix we show that the fermion bilinears $N^a = -i \bar{\Psi} T^a \Psi$ and $M = -i \bar{\Psi} \Psi$ have the same scaling dimension to all orders in the $1/N_f$ expansion.  It is simplest to do this by a direct consideration of the two-point functions of these operators, defined by
\begin{eqnarray}
C_M(x) &=& \langle (\bar{\Psi} \Psi)(x) (\bar{\Psi} \Psi)(0) \rangle \text{,} \\
C^{ab}_N(x) &=& \langle (\bar{\Psi} T^a \bar{\Psi})(x) (\bar{\Psi} T^b \Psi)(0) \rangle \text{.}
\end{eqnarray}
Precisely, we shall show that
\begin{equation}
C^{ab}_N(x) = \lambda \delta^{ab} C_M(x) \text{,}
\label{eqn:bilinear-2pt-equality}
\end{equation}
where $\lambda$ is an unimportant proportionality constant.

Diagrammatically, all contributions to both these correlators can be represented in the form shown in Fig.~\ref{fig:bilinear-2pt}.  In the case of $C^{ab}_N$ the left and right dark circles represent the flavor matrices $T^a$ and $T^b$, respectively, while for $C_M$ they are simply the identity matrix.  The important point is that in both cases these matrices are trivial in the Dirac $\gamma$-matrix space.

The shaded region represents an arbitrary combination of the elements of the large-$N_f$ perturbation theory.  These are all trivial in flavor indices, so by $SU(N_f)$ symmetry the shaded region can only give a contribution proportional to either  $\delta_{\alpha \beta} \delta_{\alpha' \beta'}$ or $\delta_{\alpha \alpha'} \delta_{\beta \beta'}$.  Let us consider these two possibilities in turn.

The first of these corresponds to the $\alpha$ fermion-line extending through the shaded area and eventually joining onto the $\beta$ line on the same side, and similarly for the $\alpha'$ and $\beta'$ lines.  So each of the external dark circles lives on a separate closed fermion loop, which is decorated with photon lines inside the shaded area.  However, because the fermion bilinears are trivial in the Dirac space, it is easy to see that these fermion loops necessarily involve a trace over an odd number of $\gamma$ matrices, which vanishes.  Therefore there are no contributions of this form.

Next we consider the second possibility, where the $\alpha$ fermion-line extends into the shaded area and emerges on the other side to join the $\alpha'$ line, and similarly for the $\beta$ and $\beta'$ fermion lines.  In this case the two dark circles reside on the same closed fermion loop, which will involve an even number of $\gamma$ matrices and can be nonzero.  Every diagram contributing to either correlation function will be of this form.  Furthermore, for every diagram contributing to $C_M$, there is a unique diagram contributing to $C^{ab}_N$ that differs only by the structure of the trace over this one-fermion loop.  In the case of $C_M$ this trace will have the form $\operatorname{Tr}(\Gamma_1 \Gamma_2)$, where $\Gamma_i$ is some matrix trivial in the flavor space.  For $C^{ab}_N$ we have instead
\begin{equation}
\operatorname{Tr}(T^a \Gamma_1 T^b \Gamma^2) = \operatorname{Tr}(T^a T^b) \operatorname{Tr}(\Gamma^1 \Gamma^2) = \lambda \delta^{ab} \operatorname{Tr}(\Gamma^1 \Gamma^2) \text{,}
\end{equation}
where $\lambda$ is a constant that is chosen once and for all by fixing a normalization for the $SU(N_f)$ generators.  This means that Eq.~(\ref{eqn:bilinear-2pt-equality}) holds diagram-by-diagram, so $N^a$ and $M$ have the same dimension to all orders in $1/N_f$.

\bibliography{su4}

\end{document}